\documentclass[12pt]{article}
\usepackage{savesym,dsfont,ytableau}
\usepackage{cite}
\usepackage{picture}
\usepackage{wasysym}
\usepackage{tikz, tikz-cd}
\usepackage{amscd}
\usepackage{graphicx}
\usepackage{pifont}
\usepackage{float}
\usepackage{subfig}
\usepackage[all]{xy}
\usepackage{multicol}
\usepackage{amsfonts}
\usepackage{picture}
\usepackage{amssymb}
\savesymbol{iint}
\savesymbol{iiint}
\usepackage{amsmath}
\restoresymbol{TXF}{iint}
\restoresymbol{TXF}{iiint}
\usepackage{color}
\usepackage{clay}
\usepackage{hyperref}
\usepackage{slashed}
\hypersetup{colorlinks=true}
\hypersetup{linkcolor=black}
\hypersetup{citecolor=black}
\hypersetup{urlcolor=black}
\usepackage{setspace}
\usepackage{multirow}
\usepackage{wrapfig}
\usepackage{verbatim}
\usepackage{accents}
\numberwithin{equation}{section}

\newcommand*{\dt}[1]{
  \accentset{\mbox{\large.}}{#1}}
\def\dot{\dt}

\begin{document}

\date{April, 2017}

\institution{IAS}{\centerline{${}^{1}$School of Natural Sciences, Institute for Advanced Study, Princeton, NJ, USA}}
\institution{PI}{\centerline{${}^{2}$Perimeter Institute for Theoretical Physics, Waterloo, Ontario, Canada N2L 2Y5}}

\title{Surface Defects and Chiral Algebras}

\authors{Clay C\'{o}rdova\worksat{\IAS}\footnote{e-mail: {\tt clay.cordova@gmail.com}}, Davide Gaiotto\worksat{\PI}\footnote{e-mail: {\tt dgaiotto@perimeterinstitute.ca}},  and Shu-Heng Shao\worksat{\IAS}\footnote{e-mail: {\tt shuhengshao@gmail.com}} }

\abstract{We investigate superconformal surface defects in four-dimensional $\mathcal{N}=2$ superconformal theories.  Each such defect gives rise to a module of the associated chiral algebra and the surface defect Schur index is the character of this module.  Various natural chiral algebra operations such as Drinfeld-Sokolov reduction and spectral flow can be interpreted as constructions involving four-dimensional surface defects.  We compute the index of these defects in the free hypermultiplet theory and Argyres-Douglas theories, using both infrared techniques involving BPS states, as well as renormalization group flows onto Higgs branches. In each case we find perfect agreement with the predicted characters.}

\maketitle

\setcounter{tocdepth}{3}
\tableofcontents

\section{Introduction}

In this paper we discuss four-dimensional $\mathcal{N}=2$ superconformal field theories coupled to conformally invariant two-dimensional $(2,2)$ surface defects $\mathbb{S}$. We investigate the defect Schur index
\begin{equation}
\mathcal{I}_{\mathbb{S}}(q)=\sum_{\mathcal{O}_{2d-4d}}\left[e^{2\pi iR}q^{R-M_{\perp}}\right]~,\label{SSchurdefintro}
\end{equation}
where in the above $R$ denotes the  $SU(2)$ $R$-charge and $M_{\perp}$ rotations transverse to the defect.  This index counts operators in the presence of the defect, which are simultaneously chiral with respect to both the left and right $2d$ supersymmetry algebras.  

Our main results are computations of these defect indices for the free hypermultiplet theory and strongly-coupled Argyres-Douglas CFTs, using both renormalization group flows along Higgs branches, as well as recent Coulomb branch formulas that express the defect indices in terms of $2d$-$4d$ BPS particles.  We also explain how our results agree with the general relationship between surface defect Schur indices and chiral algebra characters. 

\subsection{Surface Defects and Chiral Algebras}

The Schur index $\mathcal{I}(q)$ introduced in \cite{Kinney:2005ej, Gadde:2011ik, Gadde:2011uv} is a specialization of the superconformal index which counts quarter-BPS local operators in superconformal field theories.  This specialization is particularly interesting because of its remarkable connection to disparate areas of mathematical physics, including topological field theory \cite{Gadde:2009kb, Gadde:2011ik}, vertex operator algebras \cite{Beem:2013sza}, and BPS wall-crossing phenomena \cite{Iqbal:2012xm, Cordova:2015nma}.

The Schur index $\mathcal{I}(q)$ may be enriched by considering the superconformal field theory in the presence of superconformal defects, such as lines, surfaces, or boundary conditions.  These defects are useful probes of dualities and allow us to explore the possible phases of gauge theories.  The defect generalization of the Schur index then counts supersymmetric local operators bound to the defect.  Our focus in this work is on surface defects $\mathbb{S}$ that preserve $(2,2)$ superconformal symmetry.  A review of many of the properties of these defects is given in \cite{Gukov:2014gja}.  

A powerful organizing principle for general Schur indices was introduced in \cite{Beem:2013sza} and further developed in \cite{Beem:2014rza, Lemos:2014lua, Liendo:2015ofa, Lemos:2015orc, arakawa2015joseph, Nishinaka:2016hbw, Buican:2016arp, Arakawa:2016hkg, Bonetti:2016nma, beem}.  In those works, it was deduced on general grounds that for a conformal field theory, the local operators contributing to $\mathcal{I}(q)$ form a two-dimensional non-unitary chiral algebra.  In particular it follows from this analysis that the Schur index is the vacuum character of this chiral algebra.  

The chiral algebra perspective remains useful in the presence of a surface defect $\mathbb{S}$.  As we argue in section \ref{sec:chiralalg}, the defect local operators contributing to the Schur index form a module over the original chiral algebra.  Therefore the defect Schur index $\mathcal{I}_{\mathbb{S}}(q)$ is the character of some non-trivial module of the chiral algebra.  These conclusions have also been obtained in \cite{BPR}.  As we further discuss, many natural operations at the level of chiral algebras may be interpreted as constructions on four-dimensional field theories in the presence of surface defects.  

One simple observation concerns the four-dimensional meaning of spectral flow of chiral algebra modules.  Given 
a surface defect $\mathbb{S}$ in a theory with a global symmetry $U(1)$,  we may ``twist" this defect into a one-parameter family $\mathbb{S}^{(\alpha)}$.  The modules of  $\mathbb{S}$ and $\mathbb{S}^{(\alpha)}$ are then related by spectral flow by $\alpha$ units.

Another useful point concerns the behavior of the chiral algebra along Higgs branch renormalization group flows.  Let $\mathcal{T}_{UV}$ and $\mathcal{T}_{IR}$ be two theories related by a flow along the Higgs branch of $\mathcal{T}_{UV}$ parameterized by a nilpotent vev of a flavor moment map operator.  
Then, as discussed in \cite{Beem:2014rza}, the chiral algebras of $\mathcal{T}_{UV}$ and $\mathcal{T}_{IR}$ appear to be related by quantum Drinfeld-Sokolov reduction.   
This agrees with the statement that the characters, i.e. Schur indices, are related by a simple residue prescription \cite{Gaiotto:2012xa}.  Expanding on these results, one may naturally consider a class of vortex surface defects.  We claim that the chiral algebra modules associated to these defects are obtained by performing Drinfeld-Sokolov reduction on the spectral flow of the vacuum module of $\mathcal{T}_{UV}.$  In particular this agrees with the simple residue prescription for evaluating the defect Schur index \cite{Gaiotto:2012xa}.  We review this method in section \ref{sec:higgs}.

Together these techniques allow us to obtain a precise description of the chiral algebra modules associated to many surface defects. 

\subsection{The Hypermultiplet and Argyres-Douglas Theories}

After reviewing the basic properties of the surface defect Schur index and its interpretation via chiral algebras we proceed to examples.

All the theories we consider in our calculations are class $S$ theories of type $A_1$. The class $S$ definition equips these theories with an infinite family of  surface defects $\mathbb{S}_r$, labelled by a non-zero weight of $SU(2)$, i.e. a positive integer \cite{Hanany:1997vm, Gaiotto:2009fs, Gaiotto:2011tf}. All these defects also admit a uniform four-dimensional description as vortex defects \cite{Gaiotto:2012xa}. The simplest such defect, sometimes called the \textit{canonical surface defect} $\mathbb{S}_1$ (or simply  $\mathbb{S}$) also has a rather well-understood spectrum of $2d$-$4d$ BPS particles. This makes many calculations possible.  

We first turn to a detailed investigation of the surface defect $\mathbb{S}$ in the free hypermultiplet theory \cite{Gaiotto:2011tf}.  Despite the fact that the bulk theory is free, the defect $\mathbb{S}$ is in fact strongly-coupled and poorly understood.  

We use two different techniques to evaluate the index $\mathcal{I}_{\mathbb{S}}(q)$.  First, we use the infrared formulas of \cite{Cordova:2017ohl} (reviewed in section \ref{sec:2d4dBPS}) to find a formula for the defect Schur index using the $2d$-$4d$ BPS spectrum.  We compare our results with the Higgsing procedure \cite{Gaiotto:2012xa}. Both methods give the same answer, with some important subtleties.\footnote{The answer has to be treated with care due to an infinite tower of operators 
contributing an overall factor of $\sum_{n = - \infty}^\infty z^n$ to the Schur index, where $z$ is a flavor fugacity. Such infinite sums tend to give zero when the indices are manipulated as actual rational functions of the fugacities rather than generating functions.} In Appendix \ref{sec:fullhyper} we clarify these subtleties by computing the full three-variable superconformal index for the vortex defects of the hypermultiplet using the Higgsing procedure.  

We further explore the interpretation of defect local operators as a module for the bulk chiral algebra. The hypermultiplet is associated to the $\beta \gamma$ chiral algebra \cite{Beem:2013sza} and we describe the corresponding module. We also briefly discuss higher $\mathbb{S}_r$ defects. 

We then move on in section \ref{sec:AD} to consider applications to surface defects in Argyres-Douglas theories \cite{Argyres:1995jj, Argyres:1995xn}.  These theories are an ideal set of examples to illustrate the power of the infrared formula of \cite{Cordova:2017ohl} as well as the unifying chiral algebra interpretation of the resulting indices.  Indeed, the BPS spectra of Argyres-Douglas theories have been computed both with and without defects \cite{Shapere:1999xr, Gaiotto:2009hg, Gaiotto:2010be, Cecotti:2010fi, Alim:2011ae, Alim:2011kw, Maruyoshi:2013fwa, Cordova:2013bza}. Moreover the associated chiral algebras have been identified in \cite{Beem:2013sza, Beem:2014zpa, Buican:2015ina, Cordova:2015nma, Xie:2016evu, Creutzig:2017qyf, beem} and further studied in \cite{Song:2015wta, Cecotti:2015lab, Cordova:2016uwk, Fredrickson:2017yka}.  Thus, although these theories have no simple Lagrangian formulation we may still apply our technology.  We may also compare our results to the recently proposed general superconformal indices of Argyres-Douglas theories \cite{Buican:2015tda, Song:2015wta, Maruyoshi:2016tqk, Maruyoshi:2016aim, Agarwal:2016pjo}

We carry out explicit calculations for the canonical surface defects $\mathbb{S}$ of $A_{n}$ type Argyres-Douglas theories which have natural constructions in M-theory,
both using BPS spectra and Higgsing.  We find that for $A_{2n}$ our results reproduce the character of the primary $\Phi_{1,2}$ in the $(2,2n+3)$ Virasoro minimal model
\begin{equation}
\mathcal{I}_\mathbb{S}(q) = \chi^{(2,2n+3)}_{(1,2)}(q)~,
\end{equation}
for the canonical surface defect and characters of other primary fields $\Phi_{1,k+1}$ for general canonical surface defects $\mathbb{S}_k$ of not too large $k$. 
 
Meanwhile for $A_{2n+1}$ our calculations reproduce the non-vacuum characters of the $W^{(2)}_{n+1}$ chiral algebra (a certain Drinfeld-Sokolov reduction of the $SU(n+1)$ Kac-Moody algebra).  

\section{Computational Methods for $2d$-$4d$ Schur Indices }
\label{sec:indicesgen}

In this section we briefly review $2d$-$4d$ Schur indices, and various approaches that may be used to calculate them.  

A $(2,2)$ conformal surface defect $\mathbb{S}$ in a $4d$ $\mathcal{N}=2$ conformal field theory preserves the following subalgebra of the the bulk superconformal algebra $SU(2,2|2)$:
\begin{equation}
SU(1,1|1)\times SU(1,1|1)\times U(1)_{C}\subset SU(2,2|2)~.
\end{equation}
The factors $SU(1,1|1)\times SU(1,1|1)$ are the global charges of the $(2,2)$ superconformal algebra, while $U(1)_{C}$ is the commutant of the embedding and is therefore a universal flavor symmetry enjoyed by every conformal surface defect.  In terms of bulk symmetries,
\begin{equation}\label{ccharge}
C=R-M_{\perp}~,
\end{equation}
where $R$ is the Cartan of $SU(2)_{R}$ and $M_{\perp}$ generates rotations transverse to the defect.

The defect Schur index is a generalization of (a limit of) the elliptic genus of $(2,2)$ theories to include the coupling to the $4d$ bulk.  It takes the form\footnote{Here we have chosen a slightly unusual $4d$ fermion number $F_{4d}=2R$, which is more convenient in our calculations.}
\begin{equation}
\mathcal{I}_{\mathbb{S}}(q)=\sum_{\mathcal{O}_{2d-4d}}\left[e^{2\pi iR}q^{R-M_{\perp}}\right]~.
\end{equation}
Here the sum is over operators in the presence of the defect, and the variable $q$ grades these operators by their flavor charge $C$.  As compared to the most general index, the Schur limit enjoys enhanced supersymmetry. From the $(2,2)$ point of view, it receives contributions only from operators in the (chiral, chiral) sector.

\subsection{Localization Techniques}

In the special case when both the 2$d$ theory on the surface defect $\mathbb{S}$ and the 4$d$ bulk theory have Lagrangian descriptions, the $2d$-$4d$ index can be computed straightforwardly.  Since the index is invariant under marginal deformation, we can choose to work at the zero coupling point and enumerate the operators there.  The final answer for the $2d$-$4d$ index takes the form of a finite-dimensional integral and has been explored for various  $2d$-$4d$ systems in \cite{Nakayama:2011pa,Gadde:2013ftv,Cordova:2017ohl}.

First, we consider the $4d$ Schur index without defects.  Suppose the theory is defined by gauge group $G,$ and hypermultiplets in representation $\mathbf{R}$ of $G,$ and  $\mathbf{F}$ of the flavor symmetry.   The Schur index refined by flavor fugacities $x$ is given by
\begin{align}
\mathcal{I}(q,x)=\int[du] \,P.E. \left[f^{V}(q)\chi_{G}(u)+f^{\frac{1}{2}H}(q)\chi_{\mathbf{R}}(u)\chi_{\mathbf{F}}(x)\right]~,\label{operatorSchur}
\end{align} 
where $[du]$ is the Haar measure on the maximal torus of $G$ and $\chi_{\alpha}$ are characters of the gauge and the flavor group. Here $f^V(q)$ and $f^{\frac 12 H}(q)$ are the single letter indices of a free vector multiplet and a half-hypermultiplet,
\begin{align}
f^V(q) =- {2q\over 1-q}\,,~~~~~~~f^{\frac12 H} (q) = -{q^{1\over2} \over 1-q}\,.
\end{align}
Finally, $P.E.$ is the plethystic exponential,
\begin{align}
P.E. [f(q,u,x)] = \exp\left[ \sum_{n=1}^\infty {1\over n}f(q^n,u^n,x^n)\right]\,.
\end{align}

We now consider  $2d$-$4d$ systems constructed from gauging a $2d$ flavor symmetry on the defect  $\mathbb{S}$ by $4d$ gauge fields.  Using the localization formula developed in \cite{Benini:2013nda,Benini:2013xpa,Gadde:2013ftv}, we first compute a limit of the NS-NS sector elliptic genus 
\begin{equation}
\mathcal{G}_{(c,c)}(u)=\mathrm{Tr}_{NSNS}\left[(-1)^{F_{2d}}\mathbf{q}^{L_{0}-J_{0}/2}\mathbf{\bar{q}}^{\bar{L}_{0}-\bar{J}_{0}/2}u^{K}\right]~,
\end{equation}
where $u$ is a flavor fugacity for the $2d$ flavor group $K.$  This is the limit of the genus that counts only $(c,c)$ operators, and hence depends only on flavor variables.

The $2d$-$4d$ Schur index is then a simple generalization of \eqref{operatorSchur} by inserting the contribution from the defect,
\begin{align}
\mathcal{I}_\mathbb{S}(q,x)=\int[du] \,P.E. \left[f^{V}(q)\chi_{G}(u)+f^{\frac{1}{2}H}(q)\chi_{\mathbf{R}}(u)\chi_{\mathbf{F}}(x)\right]\,\mathcal{G}_{(c,c)}(u)~.
\end{align} 
For example applications of this formula see e.g. \cite{Cordova:2017ohl}.

\subsection{The Infrared Formula}
\label{sec:2d4dBPS}

In this section we review the infrared formula of \cite{Cordova:2017ohl} for computing defect Schur indices in terms of Coulomb branch data. This formula intertwines the Cecotti-Vafa formula \cite{Cecotti:1992rm, Gaiotto:2015aoa} expressing limits of the elliptic genus in terms of $2d$ BPS soliton degeneracies with recent formulas expressing the $4d$ Schur index (and generalizations) in terms of $4d$ BPS particles \cite{Iqbal:2012xm, Cordova:2015nma, Cecotti:2015lab, Cordova:2016uwk}.

On the Coulomb branch, the bulk dynamics is that of an abelian gauge theory with gauge group $U(1)^{r}$ \cite{Seiberg:1994rs, Seiberg:1994aj} ($r$ is typically called the rank of the theory), while the defect theory is typically gapped with with $N$ vacua.  There are various BPS objects that can appear in this coupled system \cite{Gaiotto:2011tf}.  These may carry electromagnetic charges $\gamma$ valued in a lattice $\Gamma$ and as well may carry spin (denoted $n$ below).  We count them with appropriate indices.
\begin{itemize}
\item $4d$ BPS particles counted by indices $\Omega(\gamma,n)$.
\item $2d$ BPS particles counted by indices $\omega_{i}(\gamma,n)$ where $i$ indicates a $2d$ vacuum.
\item $2d$ BPS solitons  counted by indices $\mu_{ij}(\gamma,n)$ where $i\neq j$ indicates a pair of $2d$ vacua.
\end{itemize}

From these indices we build a wall-crossing operator $\mathcal{S}^{2d-4d}_{\vartheta,\vartheta+\pi}(q)$ which is an $N\times N$ matrix, whose entries are power series in a quantum torus algebra of variables $X_{\gamma}$ obeying
\begin{equation}
X_{\gamma}X_{\gamma'}=q^{\frac{1}{2}\langle \gamma, \gamma' \rangle}X_{\gamma+\gamma'}~,\label{qtalg}
\end{equation}
where $\langle \gamma, \gamma' \rangle$ is the Dirac pairing.  The wall-crossing operator encodes the part of the spectrum whose central charge phases lie in the half-space $\arg(\mathcal{Z})\in [\vartheta , \vartheta+\pi)$.   It takes the form of a phase ordered product
\begin{equation}\label{refinedwall2d}
\mathcal{S}^{2d-4d}_{\vartheta,\vartheta+\pi}(q)= :\prod_{ij,\gamma| \arg(\mathcal{Z})\in [\vartheta , \vartheta+\pi)}^{\curvearrowleft} 
S_{ij;\gamma}K_{\gamma'}^{2d} K_{\gamma'}^{4d} :~,
\end{equation}
where the normal ordering means that each factor is ordered according to increasing central charge phase.

The individual matrices in \eqref{refinedwall2d} depend on the refined indices introduced above.  Let $\delta^{i}_{j}$ indicate the $N\times N$ identity matrix, and $e_{j}^{i}$ an $N\times N$ matrix whose only non-vanishng entry is in the $i$-th row and $j$-th column.   Then, the factors are defined as follows:
\begin{itemize}
\item The $4d$ particles with gauge charge $\gamma$ contribute factors of
\begin{equation}
K^{4d}_\gamma(q;\Omega_{j}(\gamma))=\prod_{n\in \mathbb{Z}}E_{q}((-1)^{n}q^{n/2}X_{\gamma})^{(-1)^{n}\Omega_{n}(\gamma)}\delta^{i}_{j}~.
\end{equation}
Here $E_q(z)$ is the quantum dilogarithm defined as
\begin{equation}
E_{q}(z) = (-q^{\frac12} z;q)^{-1}_\infty = \prod_{i=0}^{\infty}(1+q^{i+\frac{1}{2}}z)^{-1}=\sum_{n=0}^{\infty}\frac{(-q^{\frac{1}{2}}z)^{n}}{(q)_{n}}~, \label{eqdef}
\end{equation}
and as usual $(q)_{n}=\prod_{j=1}^{n}(1-q^{j})$~.
\item The $2d$ particles in the $i$-th vacuum with gauge charge $\gamma$ contribute factors of
\begin{equation}
K^{2d}(q;X_\gamma;\omega_{i}(\gamma,n)) \equiv \sum_i \prod_{n\in \mathbb{Z}} (1- (-1)^{n}q^{\frac{n}{2}}X_{\gamma})^{-\omega_i(\gamma,n)} e_{i}^{i}~.\label{2d4dK}
\end{equation}
\item The 2d solitons with topological charge $ij$ and gauge charge $\gamma$ contribute factors of
\begin{equation}
S_{ij}(q;X_\gamma, \mu_{ij}(\gamma,k)) \equiv \delta^i_j -\sum_{k\in \mathbb{Z}} \mu_{ij}(\gamma,k) (-1)^{k}q^{\frac{k}{2}} X_{\gamma} e_{j}^{i}~. \label{2d4dS}
\end{equation}

\end{itemize}
Note that the $2d$ and $4d$ particles carrying the same gauge charge $\gamma$ appear at the same phase, hence we often group their contributions together by writing 
\begin{equation}
K_\gamma \equiv K^{4d}(q;X_\gamma;\Omega(\gamma,n)) K^{2d}(q;X_\gamma;\omega_{i}(\gamma,n))~.
\end{equation}

With these preliminaries, we may now state the conjectured infrared formula of \cite{Cordova:2017ohl} for surface defect Schur 
indices.   It reads
\begin{equation}\label{IRsurfaceform}
\mathcal{I}_{\mathbb{S}}(q)=(q)_{\infty}^{2r}~\mathrm{Tr}\left[\mathcal{S}^{2d-4d}_{\vartheta, \vartheta+\pi}(q)\mathcal{S}^{2d-4d}_{\vartheta+\pi, \vartheta+2\pi}(q)\right]~,
\end{equation}
where on the right-hand side the trace operation is the ordinary trace on the $N\times N$ matrix (arising from the defect vacua) as well as a trace operation on the quantum torus algebra.  Specifically, we write the $i$-$j$-th entry of the wall-crossing operator as a series in torus algebra variables
\begin{equation}
\mathcal{S}^{2d-4d}_{ij, \vartheta, \vartheta+\pi}(q)=\sum_{\gamma}s_{ij}^{\gamma}X_{\gamma}~.
\end{equation}
Then, 
\begin{equation}
\mathrm{Tr}\left[\mathcal{S}^{2d-4d}_{\vartheta, \vartheta+\pi}(q)\mathcal{S}^{2d-4d}_{\vartheta+\pi, \vartheta+2\pi}(q)\right]=\sum_{\gamma}s_{ij}^{\gamma}s_{ji}^{-\gamma}~. \label{tracedef}
\end{equation}

We can also obtain the dependence on flavor fugacities $x$ in the index by identifying these variables with the commuting elements in the torus algebra and hence summing only over non-vanishing electromagnetic charges in \eqref{tracedef}.  

\subsection{Indices From Higgsing}
\label{sec:higgs}

Another useful way to construct surface defects and their indices is to use renormalization group flows along Higgs branches  \cite{Gaiotto:2012xa}.  Suppose we are given a UV theory $\mathcal{T}_{UV}$ with a flavor symmetry $U(1)_{f}$.  We consider a flow on the Higgs branch of $\mathcal{T}_{UV}$ along a direction which spontaneously breaks both $SU(2)_R$ and $U(1)_f$ but preserves a diagonal combination of their Cartans. In the infrared we find a target theory of interest $\mathcal{T}_{IR}$, together with free hypermultiplets that describe directions along the Higgs branch of $\mathcal{T}_{UV}$ that do not belong to the Higgs branch of $\mathcal{T}_{IR}$. 

Note that such free hypermultiplets may or may not have non-standard quantum numbers under the $SU(2)_{R}$ symmetry of the infrared theory.  For instance the dilaton multiplet, which is universally present in such a flow, has a singlet  parameterizing radial motion along the Higgs branch, as well as a triplet  of goldstone bosons parameterizing angular directions.   Such non-standard hypermultiplets are not included in the Higgs branch of $\mathcal{T}_{IR}.$\footnote{This is required so that $\mathcal{T}_{IR}$ admits a superconformal stress tensor multiplet \cite{CD}.}  On the other hand standard hypermultiplets may or may not be part of $\mathcal{T}_{IR}$. 

More generally instead of turning on a constant expectation value for the charged operator triggering the flow,  we can instead choose a position-dependent profile for the operator. One interesting class of profiles are holomorphic configurations of the Higgs branch operators on a two-dimensional plane in $\mathbb{R}^4$ preserving $2d$ (2,2) supersymmetry. These flow to (2,2) surface defects in the IR theory $\mathcal{T}_{IR}$ known as \textit{vortex surface defects}. The name is justified by a modified definition involving dynamical vortices in a theory where $U(1)_f$ is gauged. Later on, we will discuss a third perspective which constructs the same defects as a result of standard Higgs branch RG flow in the presence of a ``monodromy defect'' in the UV defined by a supersymmetric lump of background $U(1)_f$ flux. 

As long as the IR $R$-symmetry matches the preserved combination of the UV $R$-symmetry and $U(1)_f$, the superconformal index of the IR theory, 
including potential vortex surface defects, can be computed from the superconformal index of the UV theory \cite{Gaiotto:2012xa}. To understand the essential idea, consider the analytic properties of the superconformal index of $\mathcal{T}_{UV}$.  The UV Schur index has various poles in the $U(1)_f$ fugacity variable $x$.  For generic values of the fugacities, there are no flat directions in $S^3\times S^1$ background for the superconformal index and hence the index is finite.  At the specific values of the flavor fugacity $x$ where the index develops a pole, a flat direction opens up in the corresponding $S^3\times S^1$ background, allowing us to turn on a nonzero expectation value for the charged operator.  Therefore the residues of the UV index at these poles are expected to be related to the indices of the IR theory $\mathcal{T}_{IR}$.  More exactly, to obtain indices of $\mathcal{T}_{IR}$ from the residue, we must remove an overall prefactor accounting for the undesired hypermultiplets found in the flow.   

In particular, for RG flows in class $S$ theories of type $A_{N-1},$ \cite{Gaiotto:2012xa} deduced that the $2d$-$4d$ Schur index for  a vortex surface defect $\mathbb{S}_r$ (with vortex number $r\in\mathbb{N}$) are captured by the poles in the $U(1)_f$ fugacity $x$ at $x=q^{r+1\over 2}$
\begin{align}\label{higgsing}
\mathcal{I}_{\mathbb{S}_r}[\mathcal{T}_{IR}](q) = (-1)^a q^b N (q)_\infty^2 \,\text{Res}_{x=q^{r+1\over2}} {1\over x}\mathcal{I}[\mathcal{T}_{UV}](q)\,.
\end{align}
Here the monomial $(-1)^aq^b$ is a theory dependent normalization factor (fixed by demanding that the index start from one), with $a,b$ being theory-dependent constants. The special case of the ordinary Schur index is $r=0$ above which arises from the residue of the first pole at $x=q^{1/2}$.

Of course a given IR theory $\mathcal{T}_{IR}$ can typically be embedded into more than one UV theory, and the Higgsing procedure described above will generally give rise to different classes of vortex surface defects.  In later sections we will apply the Higgsing procedure to both Lagrangian theories and strongly coupled Argyres-Douglas theories.

\section{Chiral Algebra Interpretation of Surface Defects}
\label{sec:chiralalg}

The (2,2) surface defects in four-dimensional $\mathcal{N}=2$ conformal field theories admit a natural interpretation as modules of the associated chiral algebra.  In this section we demonstrate this crucial fact, and discuss the $4d$-$2d$ dictionary relating operations on modules and constructions involving surface defects.  For related work see \cite{BPR}.

Let us first review the situation in the absence of surface defects.  For any four-dimensional $\mathcal{N}=2$ superconformal field theory,  a bijective correspondence between the Schur operators  and the states in the vacuum module of an associated chiral algebra was established in \cite{Beem:2013sza}.  As a consequence, the Schur index equals the vacuum character of the chiral algebra.  The OPEs between the $4d$ Schur operators reduce to those of the chiral algebra after passing to the cohomology of certain linear combinations of supercharges $\mathtt{Q}_i,\mathtt{Q}_i^\dagger$ ($i=1,2$).

\subsection{Chiral Algebra Modules}

In this section we  show that a $(2,2)$ conformal surface defect \textit{transverse} to the chiral algebra plane preserves the four supercharges $\mathtt{Q}_i,\mathtt{Q}_i^\dagger$   of the chiral algebra cohomology.  The OPE between a $4d$ bulk Schur operator with a $2d$-$4d$ Schur operator can then be restricted to the $\mathtt{Q}_i$-cohomology.  This defines a chiral algebra action on the $2d$-$4d$ Schur operators.   

One immediate consequence is that the $2d$-$4d$ Schur operators form a module of the chiral algebra.  In particular, the Schur index for a $(2,2)$ superconformal surface defect  equals  the  character of a chiral algebra module.\footnote{On the other hand, a non-conformal surface defect in a $4d$ $\mathcal{N}=2$ superconformal field theory does not preserve the supercharges $G^\pm_{+1/2}, \bar G^\pm_{+1/2}$ that are used to construct  $\mathtt{Q}_i,\mathtt{Q}_i^\dagger$ for the chiral algebra cohomology.  Therefore,  one cannot define the chiral algebra cohomology  for this $2d$-$4d$ system, and the $2d$-$4d$ Schur index needs not be a character.}   This provides a powerful organizing principle for  superconformal surface defects in $4d$ $\mathcal{N}=2$ conformal theories.

Let us verify the above statement on supercharges below.  We will follow the convention of  \cite{Cordova:2017ohl} on the four-dimensional $\mathcal{N}=2$ superconformal algebra $SU(2,2|2)$.  Its maximal bosonic subgroup consists of the four-dimensional bosonic conformal group $SO(2,4)$ and the $R$-symmetry group $SU(2)_R\times U(1)_r$. 
 The nonvanishing anticommutators between the fermionic generators $\{Q^A_{~\alpha},~ \tilde Q_{A\dot \alpha},~S_A^{~\alpha}, ~\tilde S^{A\dot \alpha}\}$ are
\begin{align}
\begin{split}
&\{Q^A_{~\alpha}  ,  \tilde Q_{B\dot \beta} \} =2 \delta^A_B \sigma^\mu_{\alpha \dot \beta} P_\mu 
= \delta^A_B P_{\alpha \dot \beta}
\,,\\
&\{\tilde S^{A\dot \alpha} , S_B^{~\beta} \} =2 \delta^A_B \bar\sigma^{\mu\dot\alpha\beta} K_\mu  
=  \delta^A_B  K^{\dot \alpha \beta}
\,,\\
&\{Q^A_{~\alpha} , S_B^{~\beta} \} = {1\over 2} \delta^A_B \delta^\beta_\alpha \Delta +\delta^A_B M_\alpha^{~\beta} - \delta_\alpha^\beta R^A_{~B}\, ,\\
&\{\tilde S^{A\dot \alpha} ,\tilde Q_{B\dot \beta} \} ={1\over 2}\delta^A_B \delta^{\dot \alpha}_{\dot \beta} \Delta 
+\delta^A_B M^{\dot \alpha}_{~\dot \beta} +\delta^{\dot \alpha}_{\dot\beta} R^A_{~B}\,,
\end{split}
\end{align}
where $A,B =1,2$ are the doublet indices of  $SU(2)_R$, and $\alpha,\beta= +,-$, $\dot\alpha,\dot\beta= \dot +,\dot -$   are the doublet indices of $SU(2)_{1}\times SU(2)_2 =SO(4)_{\text{rotation}}$.  
Here $\Delta$ is the dilation generator and $M_\alpha^{~\beta}, M^{\dot\alpha}_{~\dot \beta}$ are the $SO(4)_{\text{rotation}}$ rotation generators.   $R^A_{~B}$ includes the generators of the $SU(2)_R$  and the  $U(1)_r$.

A (2,2) conformal surface defect $\mathbb{S}$  preserves an  $SU(1,1|1)\times SU(1,1|1)\times U(1)_C$ subalgebra of $SU(2,2|2)$. Here $SU(1,1|1)\times SU(1,1|1)$ is the global part of the 2$d$ (2,2) NS-NS superconformal algebra and $U(1)_C$ is the commutant of this embedding.  
 The nonzero (anti)commutators of  $SU(1,1|1)\times SU(1,1|1)$  are
\begin{align}
&[L_0, G^\pm_r] = -r G^\pm_r\,,~~~~~~~[\bar L_0,\bar G^\pm_r] = -r \bar G^\pm_r\,,\notag\\
&[J_0,G^\pm_r] =\pm G^\pm_r\,,~~~~~~~~~[\bar J_0, \bar G^\pm_r]=\pm \bar G^\pm_r\,,\notag\\
&\{ G^+_r ,G^-_s\}=  L_{r+s}  +{ r-s \over 2}J_{r+s} \,,~~~~\{ \bar G^+_r ,\bar G^-_s\}= \bar L_{r+s}  +{r-s\over2} \bar J_{r+s}\,,~~~~~r,s=\pm\frac12\,.
\end{align}

The symmetry group $SU(1,1|1)\times SU(1,1|1)\times U(1)_C$ of  a surface defect lying on the 12-plane can be embedded into the four-dimensional superconformal algebra by the following identification:
\begin{align}\label{preserve1}
G^+_{-\frac12} = Q^2_{~+}\,,~~~~~G^-_{-\frac12} = \tilde Q_{2\dot -}\,,~~~~~
\bar G^+_{-\frac12}  =Q^1_{~-}\,,~~~~~\bar G^-_{-\frac12} = \tilde Q_{1\dot +}\,,
\end{align}
and similarly for their superconformal counterparts,
\begin{align}\label{preserve2}
G^+_{+\frac12} =  \tilde S^{2\dot -}\,,~~~~~G^-_{+\frac12} =S_2^{~+}\,,~~~~~
\bar G^+_{+\frac12}  =\tilde S^{1\dot +}\,,~~~~~\bar G^-_{+\frac12} =S_1^{~-} \,.
\end{align}
The identifications between  the  bosonic generators can be found, for example, in \cite{Cordova:2017ohl}.

If we choose the chiral algebra plane to be the 34-plane (i.e. transverse to the surface defect plane), then the four supercharges $\mathtt{Q}_i$ and $\mathtt{Q}_i^\dagger$ $(i=1,2)$ that are used to construct the cohomology in \cite{Beem:2013sza} are 
\begin{align}
\begin{split}\label{ChiralAlgebraQ}
&\mathtt{Q}_1 = Q^1_{~-} + \tilde S^{2\dot -}\,,~~~\mathtt{Q}_2 =  S_1^{~-}- \tilde Q_{2\dot -}\,,\\
&\mathtt{Q}_1^\dagger = S_1^{~-}+ \tilde Q_{2\dot -}\,,~~~\mathtt{Q}_2^\dagger =Q^1_{~-} - \tilde S^{2\dot -}\,.
\end{split}
\end{align}
 Indeed as claimed above,  the chiral algebra supercharges $\mathtt{Q}_i$ and $\mathtt{Q}_i^\dagger$  are preserved by the supercharges \eqref{preserve1} and \eqref{preserve2} of a transverse surface defect.  
 
\subsection{Spectral Flows and Monodromy Defects}
\label{sec:monodromy}

The notion of surface defect can be slightly generalized to allow for a flavor twist: a co-dimension two ``twist'' defect may live at the end of a topological domain wall implementing some flavor group rotation.

This construction is often important in discussing canonical surface defect in class $S$ theories: both surface defects and certain protected bulk operators arise from co-dimension four defects in the  six-dimensional $(2,0)$ SCFTs. These defects have mild non-locality properties due to the fact that the $6d$ SCFTs are relative quantum field theories, i.e. live, strictly speaking, at the boundary of very simple seven-dimensional invertible topological field theories. As a consequence, some local operators may have a discrete monodromy around the canonical surface defects, which thus belong to a twisted sector. In particular, theories of type $A_1$ have a canonical $\mathbb{Z}_2$ flavor generator and $\mathbb{S}_{2n+1}$ belong to $\mathbb{Z}_2$ twisted sectors \cite{Gaiotto:2011tf}. 

A priori there should be no relation between the surface defects which are available in distinct twisted sectors. In practice, though, we have found that at least as far as BPS or protected data is concerned, including the chiral algebra data, there is no obstruction in continuously deforming a given defect $\mathbb{S}$ into a family $\mathbb{S}^{(\alpha)}$ of defects in a sector twisted by $\exp 2 \pi i \alpha J_f$ for some $U(1)$ flavor generator $J_f$. 

At the level of $2d$-$4d$ BPS spectra, this deformation simply shifts the angular momentum of BPS particles of charge $q$ by $\alpha q$. The twisted superpotential data of the surface defect can be left essentially unchanged. 

The protected part of the OPE of bulk operators at the defect, captured by the chiral algebra module relations,
\begin{equation}
{O}^i(z) |v\rangle_\mathbb{S} = \sum_{n\in \mathbb{Z}} \frac{1}{z^n} \left( {O}^i_{n- \Delta_{O^i}} |v\rangle_\mathbb{S} \right)\,,
\end{equation}
is deformed schematically to 
\begin{equation}
{O}^i(z) |v\rangle_\mathbb{S} = \sum_{n\in \mathbb{Z}} \frac{1}{z^{n + \alpha q^i}} \left( {O}^i_{n- \Delta_{O^i}} |v\rangle_\mathbb{S} \right)\,,
\end{equation}
for current algebra primaries ${O}^i$ with charges $q_i$.

More precisely, we implement the deformation as a spectral flow deformation. Given a chiral algebra with a $U(1)$ current subalgebra of level $k$, normalized so that the charges of operators in the algebra are integral, we can always bosonize the current as $J = - i \partial \phi$ and consider a free boson vertex operator 
\begin{equation}
V_\alpha(z)  = e^{i \alpha \phi}(z) .
\end{equation}
This defines a module which we can call ``spectral flow of the vacuum module by $\alpha$ units''.

More generally, we can take the OPE of $e^{i \alpha \phi}(z)$ and any other module for the chiral algebra 
to produce a new, spectral flowed, module. Concretely, this can be described as the image of the original module 
under the action of the exponentiated zero mode $e^{i \alpha \phi_0}$. Alternatively, we can think about the spectral flow as an 
automorphism on the chiral algebra. It acts on the current, the Virasoro generators and other 
primaries $O^i$ of $U(1)$ charge $q_i$ as
\begin{align}
\begin{split}
&O^i_r \to {O^{i}}'_r=O^{i}_{r+\alpha q^i} \,,\\
&J_n \to J_n'  =  J_n +a \alpha\delta_{n,0}\,,\\
&L_n \to L_n' = L_n +\alpha J_n +  {k\over 2} \alpha^2 \delta_{n,0}\,,
\end{split}
\end{align} 

By taking $\alpha$ to be integer we can deform any surface defect $\mathbb{S}$ to an infinite discrete family of standard defects $\mathbb{S}^{(n)}$. If $\mathbb{S}$
is the trivial defect we will call $\mathbb{S}^{(n)}$ \textit{monodromy defects}.

We can give a physical justification for this deformation operation by observing that the components of a background bulk $U(1)$ connection in the plane orthogonal to the surface defect enters the $(2,2)$ Lagrangian as the background value of a chiral multiplet, irrespectively of their dependence on the transverse directions. For example, the transverse kinetic terms of hypermultiplets arise from superpotential terms of the schematic form $X D_{\bar z} Y$. 

As a consequence, we can turn on such a background connection without breaking $(2,2)$ supersymmetry. 
If we take our connection to have a lump of $\alpha$ units of flux in the neighbourhood of the surface defect $\mathbb{S}$
and flow to the IR, we will end up with a new surface defect $\mathbb{S}^{(\alpha)}$ in a twisted sector shifted by $\alpha$. 
This manipulation affects protected quantities exactly as we desire for a monodromy defect.

\subsection{Drinfeld-Sokolov Reduction and Higgsing}
\label{sec:gends}

As discussed in section \ref{sec:higgs}, it is often the case one can find pairs of four-dimensional ${\cal N}=2$ SCFTs, $\mathcal{T}_{UV}$ and $\mathcal{T}_{IR}$ which are related by an RG flow initiated by expectation values for Higgs branch operators.  As the indices of $\mathcal{T}_{UV}$ and $\mathcal{T}_{IR}$ are related it is natural to expect that the chiral algebras are also related. 

One challenge to this idea is that the construction of the chiral algebra relies heavily on superconformal symmetry.   Since the Higgs branch flow spontaneously breaks this symmetry it is not completely obvious that there should be a standard prescription to compute the chiral algebra of $\mathcal{T}_{IR}$ from the chiral algebra of $\mathcal{T}_{UV}$. As proposed in \cite {Beem:2014rza}, there is a very special situation where a candidate prescription exists and matches the index prescription: the situation where $\mathcal{T}_{UV}$ has a non-Abelian flavor symmetry $G$ and the RG flow is triggered by the corresponding moment map operators $\mu_G$
getting an expectation value in some nilpotent direction, identified with the raising operator $t^+$ of some $\mathfrak{su}(2) \to \mathfrak{g}$ embedding, with the $\mathfrak{su}(2)$ Cartan generator $t^3$ playing the role of the spontaneously broken $U(1)_{f}$ flavor symmetry in section \ref{sec:higgs}. 

Then the corresponding chiral algebra operation is a quantum Drinfeld-Sokolov (qDS) reduction, which consists of three steps \cite{drinfeld1984lie}:
\begin{itemize}
\item The stress tensor is shifted as $T \to T - t^3 \cdot \partial J$ 
so that the WZW current $t^+ \cdot J$ corresponding to the operator getting a vev 
has scaling dimension $0$. 
\item Decompose the Lie algebra as $\mathfrak{g} = \oplus_n \mathfrak{g}_n$ according to the $t^3$ charge.\footnote{The case where the charges of $t^{3}$ are odd has additional technical complications.  See \cite{deBoer:1992sy, deBoer:1993iz} for details.} 
Select a nilpotent subalgebra $\mathfrak{n} =\oplus_{n> 1} \mathfrak{g}_n$. These are the 
currents which have acquired dimension less than or equal to zero after the shift of the stress tensor, corresponding to 
free hypermultiplets with non-standard $SU(2)_R$ charge at the bottom of the RG flow. We may also include in $\mathfrak{n}$ some  subspace of $\mathfrak{g}^*_1$, 
which is Lagrangian under the symplectic pairing $t^+ \cdot [\_, \_]$, depending on how many free hypermultiplets with standard $SU(2)_R$ charge
do we want to keep in $\mathcal{T}_{IR}$.
\item Add a collection of $bc$ ghosts is added, with $c$ valued in $\mathfrak{n}^*$. 
Take the cohomology by a standard BRST charge which sets $t^+ \cdot J =1$ in cohomology and all other currents in $\mathfrak{n}$ to $0$ in cohomology.
\end{itemize}

Notice that this prescription is usually employed on a $G$ current algebra, but it can also be employed on a general vertex operator algebra 
which has a $G$ current sub-algebra, as all operations and in particular the BRST charge only employ the Kac-Moody currents. 

We can easily extend this discussion to the vortex surface defects of $\mathcal{T}_{IR}$ obtained by flows involving position dependent Higgs branch fields from $\mathcal{T}_{UV}$.  Indeed, the qDS reduction can be implemented on modules for the chiral algebra as well leading to modules for the qDS-reduced chiral algebra.  The modules associated to the vortex surface defects can be described simply in this language: \textit{they are the qDS reduction of a spectral flow image of the vacuum module of $\mathcal{T}_{UV}.$}  Note that this proposal matches the residue computation of the index for vortex surface defects discussed in section \ref{sec:higgs}.  It also allows an useful alternative point of view on vortex defects themselves as the infrared image of monodromy defects in $\mathcal{T}_{UV}$ associated to the $U(1)_{f}$ flavor symmetry employed in the RG flow.

The best known example of qDS reduction maps a Kac-Moody $SU(2)_\kappa$ VOA to a Virasoro VOA 
with $c = 13 + 6 (\kappa + 2) + 6 (\kappa + 2)^{-1}$. For general values of $\kappa$ where the vacuum module of $SU(2)_\kappa$ has no null vectors, 
this is a particularly simple reduction. The spin $j$ Weyl modules for $SU(2)_\kappa$ 
map to degenerate modules of type $(1,2j+1)$, while the spectral flow images of the vacuum module
of $SU(2)_\kappa$ are mapped to degenerate modules of type $(n+1,1)$ and spectral flowed images of spin $j$ modules to degenerate modules of type $(n+1,2j+1)$. 

For example, the qDS reduction acts on 
the vacuum character of $SU(2)_{\kappa}$ by adding ghosts and then setting the $SU(2)$ Cartan fugacity $x_2 \to q^{\frac12}$: 
\begin{equation}
\frac{1}{(q)_\infty(q x_2^2;q)_\infty(q x_2^{-2};q)_\infty} \to \frac{(x_2^2;q)_\infty(q x_2^{-2};q)_\infty}{(q)_\infty(q x_2^2;q)_\infty(q x_2^{-2};q)_\infty}\to \frac{1-q}{(q)_\infty}\,,
\end{equation}
gives the standard Virasoro vacuum module, with no other null vectors except $L_{-1}|0\rangle$. 

If we introduce $r$ units of spectral flow before the qDS reduction, the spectral flowed vacuum module maps to a degenerate Virasoro modules of type $(r+1,1)$ with a null vector at level $r+1$:
spectral flowing $x_2 \to q^{\frac{r}{2}} x_2$, adding ghosts and then setting $x_2 \to q^{\frac12}$: 
\begin{equation}
\frac{q^{\kappa \frac{r^2}{4}}x_2^{\kappa r}(x_2^2;q)_\infty(q x_2^{-2};q)_\infty}{(q)_\infty(q^{r+1} x_2^2;q)_\infty(q^{1-r} x_2^{-2};q)_\infty}\to (-1)^r \frac{q^{\Delta_{(r+1,1)}}(1-q^{r+1})}{(q)_\infty}\,.
\end{equation}

For special rational values of $\kappa$ where the vacuum module of $SU(2)_\kappa$ is smaller,
we expect the correspondence to be somewhat modified.  We will encounter precisely such examples of DS reduction 
when we employ the Higgsing procedure relating $D_{n+3}$ and $ A_n$ Argyres-Douglas theories in section \ref{sec:AD}. 

\section{The Free Hypermultiplet}
\label{sec:freehyper}

We begin our investigation of examples with the free hypermultiplet.  The $4d$ BPS spectrum consists of a single hypermultiplet particle and its antiparticle.  Hence the charge lattice $\Gamma$ is one-dimensional and is generated  by the flavor charge $\gamma$ of the hypermultiplet.  The Coulomb branch is a point and there is no wall-crossing phenomenon.  

The story becomes more interesting when we introduce a canonical surface defect $\mathbb{S}$ into the free hypermultiplet theory.  This surface defect is actually a rather mysterious, strongly interacting object.   It may be defined through a class S construction \cite{Gaiotto:2011tf}.    We realize the hypermultiplet by two M5-branes on the complex plane with an irregular singularity at infinity.  The canonical defect is obtained placing an M2-brane at a point $z$ on the plane. This M2-brane extends along two spacetime dimensions and hence gives rise to a surface defect.

There is no known explicit Lagrangain construction of this defect.  It is expected that it is conformally invariant and breaks the $SU(2)$ flavor symmetry to a $U(1)$ subgroup. It should be most naturally viewed as a $\mathbb{Z}_2$-twisted defect, around which the free hyper is anti-periodic.
The point $z\in \mathbb{C}$ at which the defect is placed corresponds to a twisted chiral relevant deformation of dimension $1/2$, compatible with a twisted mass for the $U(1)$.  The defect has two massive vacua when either or both of these parameters are activated. 

If $z$ is sufficiently large, the mass of the solitons between the two vacua grows large and the surface defect is expect to ``simplify'' to a sum of two simpler defects with a single vacuum: 
two monodromy defects of twist $\pm \frac12$. Furthermore, multiple line defects exist which interpolate between the full surface defect and either of these simpler defects. 

Our aim will be to learn about the spectrum of chiral operators on the defect using the infrared formula for the Schur index as well as Higgsing.  To apply our formula \eqref{IRsurfaceform} we require the full spectrum of this $2d$-$4d$ system.   The spectrum depends on the defect parameter $z$ and there is wall-crossing as $z$ is varied.  Since our IR formula \eqref{IRsurfaceform} for the $2d$-$4d$ Schur index is wall-crossing invariant, we can choose to work in a chamber with the simplest BPS particle spectrum.  

As shown in \cite{Gaiotto:2011tf}, if we shift our conventions so that the surface defect is untwisted, there is a chamber where the BPS particle spectrum  consists of a single $4d$ hypermultiplet particle with (flavor) charge $\gamma$, and a single $2d$-$4d$ soliton interpolating from vacuum one to vacuum two, as well as their antiparticles.  The is also a $2d$ particle with $\omega_2(\gamma,1)=-1$ and its antiparticle with $\omega_2(-\gamma,-1)=1$. The phase order is such that $\mathcal{Z}_{12}(z)<\mathcal{Z}_{\gamma}(z)$.    The corresponding factors are:
\begin{equation}
S_{12;0} =  \begin{pmatrix}1 & -1 \cr 0 & 1\end{pmatrix} \,,\qquad \qquad S_{21;0} =  \begin{pmatrix}1 & 0 \cr 1 & 1\end{pmatrix}\,,
\end{equation}
and 
\begin{equation}
K_\gamma=  \begin{pmatrix}E_q(X_\gamma) & 0 \cr 0 & E_q(qX_\gamma)\end{pmatrix} \,,\qquad \qquad K_{-\gamma}=  \begin{pmatrix}E_q( X_{-\gamma}) & 0 \cr 0 & E_q(q^{-1}X_{-\gamma})\end{pmatrix} \,,
\end{equation}
where we incorporated the 2d contributions $1+q^{\frac12} X_\gamma$ and $(1+q^{-\frac12} X_{-\gamma})^{-1}$ into the 4d factors by a shift of their arguments. 

If we go back to the conventions where the surface defect is a twist defect, we shift of the flavor fugacity $X_\gamma \to q^{-\frac12}X_\gamma$.  
The wall-crossing factors for the canonical surface defect are then more symmetric
\begin{equation}
K_\gamma=  \begin{pmatrix}E_q(q^{-\frac12} X_\gamma) & 0 \cr 0 & E_q(q^{1\over2} X_\gamma)\end{pmatrix} \,,\qquad \qquad K_{-\gamma}=  \begin{pmatrix}E_q( q^{1\over2}X_{-\gamma}) & 0 \cr 0 & E_q(q^{-\frac12}X_{-\gamma})\end{pmatrix} \,,
\end{equation}
with $S_{12;0}$ and $S_{21;0}$ the same as above.  

The physical interpretation is very simple at large $z$: the diagonal entries of the $K_\gamma$ factors collect the modes of the bulk hypermultiplet in the presence of 
the monodromy defects of twist $\pm \frac12$, the $S$ factors contain the contributions of the solitons between these two vacua. 

Our IR formula for the $2d$-$4d$ index gives
 \begin{eqnarray}\label{IRhyper}
\mathcal{I}_{\mathbb{S}}(q,x)&=&\text{Tr} \left[S_{12;0}  K_\gamma S_{21;0}  K_{-\gamma }\right] \nonumber\\
&=  &\text{Tr} \left[ 
 \left(\begin{array}{cc} E_q(q^{-\frac12}X_\gamma) & -E_q(q^{\frac12}X_\gamma)  \\0 & E_q(q^{\frac12}X_\gamma)\end{array}\right) 
  \left(\begin{array}{cc} E_q(q^{\frac12}X_{-\gamma}) & 0\\E_q(q^{\frac12}X_{-\gamma})& E_q(q^{-\frac12}X_{-\gamma})\end{array}\right) 
 \right] \nonumber\\
&= &E_q(q^{-\frac12}x) E_q(q^{\frac12}x^{-1}) -E_q(q^{\frac12}x) E_q(q^{\frac12}x^{-1}) 
+E_q(q^{\frac12}x) E_q(q^{-\frac12}x^{-1})
\label{canonicalhyperzero}\\
&= &  E_q(q^{\frac12}x) E_q(q^{\frac12}x^{-1})  \, \left[ {1 \over  1+x}  -1 + {1\over 1+x^{-1}}\right] 
 \nonumber\\
&=& 0~,\nonumber
\end{eqnarray}
where in the above $x = \text{Tr}[X_\gamma]$ is the fugacity for the $U(1)$ flavor symmetry and is identified with the flavor variable $X_{\gamma}$ in the quantum torus algebra \eqref{qtalg}.  
In the final step above we have proceeded naively and cancelled the two summands against each other. Formally, that would mean that the Schur index decorated by the canonical surface defect vanishes.   This may mean one of two things: either the canonical defect has no protected chiral operators, including the identity, and hence likely breaks supersymmetry, or we were too hasty in our manipulations. We believe the second possibility actually occurs.
 
Indeed, our final manipulation was very suspicious: each of the two fractions in the penultimate step of our calculation really stands for 
an infinite geometric series, respectively in $x$ and $x^{-1}$, counting operators of non-zero positive or negative flavor charge and $R = M_\perp$. 
It seems too glib to resum these series and cancel them against each other. Instead, we should have probably written our answer as 
a formal Laurent series:
\begin{equation}
\mathcal{I}_{\mathbb{S}}(q,x) = E_q(q^{\frac12}x) E_q(q^{\frac12}x^{-1}) \sum_{n \in \mathbb{Z}} x^n .
\end{equation}
We will encounter a similar phenomenon when we compute the index in other ways: the answer naively vanishes, but only when we allow 
cancellations between rational terms which arise from power series with different regions of convergence.

\subsection{$\beta\gamma$-System Modules and the Canonical Surface Defect}
\label{chiralhyper}
We will now explore the chiral algebra meaning of the index found in the previous section.\footnote{We would like to thank Thomas Creutzig for an enlightening discussion on this point.}  As discussed in sections \ref{sec:chiralalg}, the Schur operators on a conformal surface defect are a module over the chiral algebra associated to the bulk $\mathcal{N}=2$ theory.  In the case at hand the associated chiral algebra is simply the $\beta\gamma$ system with weight $h_\beta=h_\gamma=\frac12$, also known as symplectic bosons \cite{Lesage:2002ch}.  The  OPE is,
\begin{equation}
\beta(w)\gamma(0)\sim \frac{1}{w}~, \hspace{.5in}\gamma(w)\beta(0)\sim- \frac{1}{w}~.
\end{equation}
Let $\beta_n$ and $\gamma_n$ be the modes of these currents,
\begin{align}
\beta(w) =  \sum_{n\in \mathbb{Z}+\nu } {\beta_n \over w^{n+\frac12}}\,,~~~~~~~
\gamma(w) = \sum_{n\in \mathbb{Z}+\nu}  {\gamma_n \over w^{n+\frac12}}\,,
\end{align}
where $\nu=0$ ($\nu=\frac12$) for a $\mathbb{Z}_2$-twisted (untwisted) module.  The algebra of the modes is
\begin{align}\label{modealgebra}
[\beta_{m}, \gamma_{n}]=\delta_{m,-n}\,.
\end{align}
We can define a weight-1 current $J(w)$ descended from the   4$d$ $U(1)$ flavor symmetry
\begin{align}
J(w) =  - : \gamma\beta: (w)\,,
\end{align}
under which $\beta$ and $\gamma$ carry $+1$ and $-1$ charge, respectively.

The untwisted vacuum module is very simple: it is generated by a vacuum which is annihilated by all positive modes of $\beta$ and $\gamma$. 
The  $\mathbb{Z}_2$-twisted (aka Ramond) modules are more subtle, because of the presence of the zero modes $\beta_0$ and $\gamma_0$, which form an Heisenberg algebra. 
Simple highest weight twisted modules for the symplectic boson can be induced from any module for this Heisenberg algebra.

Spectral flow by $\pm \frac12$ units of the vacuum module produce twisted modules generated by 
Heisemberg modules respectively of the form $\beta_0^n |+\rangle$ with $\gamma_0 |+\rangle=0$ 
and $\gamma_0^n |-\rangle$ with $\beta_0 |-\rangle=0$. 

Physically, that means a defect OPE where one of the two complex fields in the hypermultiplet diverges as $w^{-\frac12}$ and the other 
goes to zero as $w^{\frac12}$ as a function of the transverse coordinate $w$. The corresponding characters are
\begin{equation}
\mathcal{I}_{\pm}(q,x) = E_q(q^{\frac12}x) E_q(q^{\frac12}x^{-1}) x^{\pm \frac12} \sum_{n =0}^\infty x^{\pm n}\,,
\end{equation}
with a semi-infinite sum associated to either $\beta_0$ or $\gamma_0$ zero modes. 

The direct sum of these two modules has an interesting deformation: we may set $\gamma_0 |+\rangle=c |-\rangle$ and
$\beta_0 |-\rangle=c |+\rangle$. This produces a ``bilateral'' module with a formal character 
\begin{equation}
E_q(q^{\frac12}x) E_q(q^{\frac12}x^{-1}) x^{\frac12- c^2} \sum_{n \in \mathbb{Z}}^\infty x^{ n}\,.
\end{equation}
Up to the overall power of $x$, which is normally neglected in the index, this agrees with $\mathcal{I}_{\mathbb{S}}(q,x)$.

Notice that the deformation shifts the current eigenvalue of $|\pm \rangle$ to $\pm \frac12 - c^2$. It  is useful to denote $\lambda = \frac12-c^2$ and the module as $R_\lambda$. The actual module depends on $\lambda$ up to integer shifts and degenerates when $\lambda$ is half-integral. 

Physically, we expect that the surface defect supports an infinite series of chiral operators, which we are tempted to denote as $e^{n \Phi}$ with $\Phi$ being some chiral field valued on a cylinder, so that the two complex scalars in the hypermultiplet have a defect OPE
\begin{equation}
X(w) \sim c e^{\Phi} w^{-\frac12} + O(w^{\frac12}) \,,\qquad \qquad Y(w) \sim c e^{- \Phi} w^{-\frac12} + O(w^{\frac12})~.
\end{equation}

\subsection{The Full Defect Index from Higgsing}

We can gain more information about this defect using the Higgsing procedure described in section \ref{sec:higgs}.  The free hypermultiplet can be obtained by moving onto the Higgs branch of the $D_4$  Argyres-Douglas theory whose Schur index is explicitly known \cite{Buican:2015ina,Cordova:2015nma}.  The canonical surface defect of the hypermultiplet is a vortex defect from this point of view and hence its Schur index can be obtained by taking an appropriate residue. We refer for further details to section \ref{sec:AD} where we discuss Higgsing of general $D_{n+4}$ Argyres-Douglas theories.  

The  Schur index for the charge $r$ vortex defect obtained from Higgsing the $D_4$ theory is, up to an overall power of $q$, (see \eqref{higgsDeven})
\begin{equation}
\mathcal{I}_{\mathbb{S}_r}(x,q) = \frac{1}{(q)_\infty^2} \sum_{k = 0}^\infty  \left[F_{4k-r+1}-F_{4k+r+3}\right]
\end{equation}
where
\begin{equation}
F_s = \frac{q^{\frac{s^2}{8}} \left(1-\eta^2 q^s\right)}{\left(1-\eta x^{-1}q^{\frac{s}{2}}\right) \left(1-\eta x q^{\frac{s}{2}}\right)}
\end{equation}
and we included a convergence factor $\eta$ to keep track unambiguously of how the denominators 
were expanded out in the original generating function: one should first expand in positive powers of $\eta$ and only 
then set $\eta \to 1$. 

When $r=0$ all denominator fugacities have positive powers of $q$ and the meaning of this expression is 
unambiguous. It expands out to match the standard free hypermultiplet index. 

For $r=1$ the sum telescopes to a simple answer involving a single term 
\begin{equation}
\frac{1}{(q)_\infty^2}F_{0} = \frac{1}{(q)_\infty^2}\frac{1- \eta^2}{(1- \eta x)(1- \eta x^{-1})} = \frac{1}{(q)_\infty^2} \sum_{n \in \mathbb{Z}} \eta^{|n|} x^n
\end{equation}
This is the same answer as we obtained before, because the infinite sum behaves as a delta function: 
\begin{equation}
x^k \sum_{n \in \mathbb{Z}} x^n =  \sum_{n \in \mathbb{Z}} x^n
\end{equation}
and thus the infinite products $(q)_\infty^2$ match the expected $E_q(q^{\frac12}x) E_q(q^{\frac12}x^{-1})$.

For $r=2$, the sum can be expressed in terms of the $r=0$ answer: 
\begin{equation}
\mathcal{I}_{\mathbb{S}_2} = \frac{1}{(q)_\infty^2}\left(F_1 + F_{-1}\right)- \mathcal{I}_{\mathbb{S}_0}
\end{equation}
The $F_1$ and $F_{-1}$ terms can be suggestively identified with the sums of indices for  
monodromy defects of parameters $1$ and $0$, and $0$ and $-1$ respectively. 
This suggests that the Schur index of $\mathbb{S}_2$ should be thought of as the sum of the Schur indices of three
monodromy defects, of monodromy parameters $1$, $0$ and $-1$ respectively.  The corresponding module should be some 
interesting deformation of that sum of spectral flowed vacuum modules.

The pattern continues for higher $r$: the regulated Schur indices of $\mathbb{S}_r$ take the form of a sum of 
the Schur indices of $r+1$ monodromy defects, of monodromy parameters $-\frac{r}{2}$, $-\frac{r}{2}+1$, $\cdots$, $-\frac{r}{2}$.
This agrees very nicely with the class S description of the defects: all these defects admit a twisted F-term 
deformation parameter $z$ of scaling dimension $\frac12$ which triggers an RG flow to 
such a sum of $r+1$ independent monodromy defects. 

There are two ways we can go beyond the Schur index calculation and probe aspects of the surface defects 
which tease out the physical difference between $\mathbb{S}_r$ and a direct sum of monodromy defects: 
we can either compute the full superconformal index or analyze in more detail the chiral algebra modules. 

\subsubsection{Full Index Calculations}
The full superconformal index of the $D_4$  Argyres-Douglas theory has been computed in \cite{Agarwal:2016pjo}.  
Using this result together with Higgsing we present in Appendix \ref{sec:fullhyper} the full superconformal index of the canonical surface defect of the free hypermultiplet.  
Amusingly, the answer is still given as a sum of $r+1$ terms, though we do not expect each individual term to 
have a separate physical meaning. 

We leave a full discussion of the full index to future work. It would be nice to identify the operators dual to 
the expected chiral and twisted chiral deformations.  

We focus here on the Macdonald limit of this index.  From the general formula \eqref{fullhyperindapend} we specialize by taking $p=0$.  This limit is significant because it receives contributions from the same set of operators as the Schur index which is obtained when $t=q$.  The explicit expression for the canonical defect is
\begin{equation}\label{macdonald}
\mathcal{I}_{\mathbb{S}}(q,a,\epsilon)=
{1- \epsilon^{2}\over  (a \epsilon ;q)_\infty (a^{-1}\epsilon;q)_\infty }~,~~~~~\epsilon\equiv \sqrt{t\over q}\,.
\end{equation}
As compared to the Schur limit, the index is now completely well-defined.  

Interestingly, the Macdonald index \eqref{macdonald} also has a natural interpretation in terms of the twisted module $R_\lambda$ introduced in section \ref{chiralhyper}.  As before, the fugacity $a$ measures the charge associated to the current $J(w)=-:\gamma\beta:$ of the symplectic bosons.    The chiral algebra interpretation of the additional fugacity $\epsilon$, on the other hand, requires  further explanation.  Let us pick a reference state, say, $|m=0\rangle$, in the $R_\lambda$ module at level zero.  We can then construct the rest of the states in the $R_\lambda$ module by acting $\beta_{n\le0},\gamma_{n\le0}$ on $|m=0\rangle$.  Given a state in $R_\lambda$, the fugacity $\epsilon$ measures the  minimal number of modes $\beta_{n\le0},\gamma_{n\le0}$ that is required to construct this state from the reference state $|m=0\rangle$.  

With this chiral algebra definition of the $\epsilon$ fugacity, we see immediately that the Macdonald index \eqref{macdonald} is the character of $R_\lambda$ (for generic $\lambda$).  The terms $(a\epsilon;q)_\infty$ and  $(a^{-1}\epsilon;q)_\infty$ in the denominator come from the  modes  $\beta_{n\le0}$ and $\gamma_{n\le0}$, respectively.  The term $1-\epsilon^2$ in the numerator comes from the relation in the module that $\beta_0\gamma_0|\psi\rangle$ equals to $|\psi\rangle$ itself up to a nonzero multiplicative constant.  The chiral algebra interpretation of the Macdonald index of the more general Argyres-Douglas theories has been explored in \cite{Song:2016yfd}.

\subsubsection{Quantum Drinfeld-Sokolov Reduction}
According to the general discussion of section \ref{sec:gends}, it should be possible to build explicitly the modules for all the $\mathbb{S}_r$ defects by qDS 
reduction of the spectral flow of the vacuum module of the $D_4$ theory. 
 
The $D_4$ theory chiral algebra is a WZW model $SU(3)_{-\frac32}$. The Higgs branch RG flow corresponds to doing a qDS reduction based on an $SU(2)_{-\frac32}$ subalgebra. At general level, the qDS reduction of an $SU(3)_\kappa$ Kac-Moody algebra based on an $SU(2)_\kappa$ sub-algebra would give a vertex algebra generated by the following currents:
\begin{itemize}
\item A $U(1)$ current of level $\frac{2}{3} \kappa$, arising from the $U(1)$ generator in $SU(3)$ which commutes with the $SU(2)$ subgroup. 
\item A Virasoro generator of central charge $c= 23 - \frac{24}{\kappa + 3}- 6(\kappa + 3)$, all which is left of the $SU(2)$ currents. 
\item Two dimension $\frac32$ bosonic fields, arising from the two $SU(3)$ generators of charge $\frac12$ under the Cartan of $SU(2)$.
\item Two dimension $\frac12$ bosonic fields, arising from the two $SU(3)$ generators of charge $-\frac12$ under the Cartan of $SU(2)$.
\end{itemize}

The two dimension $\frac12$ fields are free symplectic bosons which are usually stripped off, leaving the Bershadsky-Polyakov algebra $W_3^{(2)}$. 
At the special level we are interested in, though, the qDS reduction collapses to the symplectic bosons and all other currents are functions of the symplectic bosons 
themselves. 

Notice that the symplectic bosons $\beta$, $\gamma$ arise directly from the two corresponding currents $J^{+1}$, $J^{+2}$, as the BRST current is just $c (J^{++}-1)$ 
and thus the BRST charge commutes with these currents. The OPE of these two currents is 
\begin{equation}
J^{+1}(w)J^{+2}(0) \sim \frac{J^{++}(0)}{w}
\end{equation}
but $J^{++}$ is BRST equivalent to $1$ and the OPE reduces to the free symplectic boson OPE. 

In order to produce ``vortex'' modules associated to vortex defects, we start from a spectral flow module associated to the Cartan of $SU(2)$, 
with the property that the current $J^{++}$ which will be set to $1$ by the qDS reduction has a pole of order $r$ while the 
current $J^{--}$ has a zero of order $r$. Inspection of the character shows clearly that the highest weight 
vector of the vortex module is the combination of the spectral flowed image of the vacuum for both the current 
algebra and the auxiliary ghost system, so that the $b$ ghost also has a pole of order $r$ and the $c$ ghost 
has a zero of order $r$. 

The symplectic bosons have both poles of order $\frac{r}{2}$ in the spectral flowed sector. 
Looking at the character of the qDS reduction, organized by $U(1)_{-1}$ charge sector, 
\begin{equation}
\chi_r[qDS_2 \circ SU(3)_{-\frac32}] = (-1)^r \sum_{n = - \infty}^{\infty} \frac{x_1^n q^{- \frac{n^2}{2}}}{(q)_\infty} \sum_{k=0}^\infty  \frac{(1-q^{(2k + 1 + |n|)(r+1)}) q^{\frac12 (2 k + |n|-\frac{r}{2})(2 k + |n|+1-\frac{r}{2})}}{(q)_\infty}
\end{equation}
we recognize for each $n$ the leading term in the character $q^{\frac{r(r-2)}{8} +|n| \frac{1-r}{2}}$ 
corresponding to states of the form $\beta_{\frac{r-1}{2}}^n |r\rangle$ and $\gamma_{\frac{r-1}{2}}^n |r\rangle$.

We leave a determination of the full structure of the vortex modules to future work. For $r=1$, the very fact that the $U(1)_{-1}$ charges are integral 
strongly suggests that we are looking at the symmetric Ramond module $R_{\lambda = 0}$.

\section{Argyres-Douglas Theories}
\label{sec:AD}

In this section we discuss canonical surface defects of the Argyres-Douglas theories.  Our aim is to compute surface defect Schur indices and to describe the associated chiral algebra modules.  We apply both the IR formula for the $2d$-$4d$ index as well as the Higgsing procedure. We will consider Argyres-Douglas theories whose BPS quivers are Dynkin diagrams of simply-laced Lie algebras $G$.\footnote{These theories are also known as the $(A_1,G)$ Argyres-Douglas theories in the terminology of \cite{Cecotti:2010fi}.}  We will demonstrate that the surface defect indices we obtain equal the characters of the associated chiral algebra.  The simplest examples are the $A_{2n}$ Argyres-Douglas theory, where the surface defect indices of different vortex numbers reproduce all and only the $n+1$ characters of the $(2,2n+3)$ Virasoro minimal model.   In the case when some flavor symmetries are broken by the defect (e.g. the canonical surface defect of the $A_3$ Argyres-Douglas theory), the surface defect index  turns out to be a  twisted character of the chiral algebra.

\subsection{$A_{2n}:$  $(2,2n+3)$ Virasoro Minimal Models}

\subsubsection{Indices from BPS States: The $A_2$ Theory}

As a warm up, let us consider the simplest nontrivial Argyres-Douglas theory whose BPS quiver is the $A_2$ Dynkin diagram.  It has a complex one-dimensional Coulomb branch and no flavor symmetry.  The associated chiral algebra is the (2,5) Virasoro minimal model with central charge $c=-22/5$, a.k.a. the Lee-Yang model.  The (2,5) Virasoro minimal model contains one non-vacuum module whose primary $\Phi_{1,2}$ has conformal dimension $h=-1/5$.  Let us first compute the canonical surface defect index using our IR formula from the BPS states.

  In one chamber there are two pure $4d$ BPS particles $\gamma^1$, $\gamma^2$ with Dirac pairing $\langle \gamma^1,\gamma^2\rangle=+1$.  In the presence of the canonical surface defect, there is a subchamber where in addition we have  one $C$ charge neutral $2d$ soliton interpolating the two vacua with vanishing 4$d$ charge and degeneracy $\mu_{12}(0,0)=1$, and one 2$d$ BPS particle living in the second vacuum with $\omega_2(\gamma^1,1) = -1$  \cite{Gaiotto:2011tf}.  Their central charge phases in increasing order are
\begin{align}
\gamma_{12}\,, \gamma^1\,,\gamma^2\,.
\end{align}
where $\gamma^1$ collectively stands for the 2$d$ particle living in the second vacuum and the 4$d$ particle, both with charge $\gamma^1$.  
Their corresponding wall-crossing factors are\begin{align}
\begin{split}
&S_{12;0} = \left(\begin{array}{cc}1 & -1 \\0 & 1\end{array}\right)\,,~~~~~~
K^{2d}_{\gamma^1} = \left(\begin{array}{cc}1 & 0 \\0 & 1+q^{1\over2} X_{\gamma^1}\end{array}\right)\,,~~~~~~K^{4d}_{\gamma^1}   =  \left(\begin{array}{cc}E_q(X_{\gamma^1} )& 0 \\0 & E_q(X_{\gamma^1})\end{array}\right)\,,~~~~~~\\
&S_{21;0} = \left(\begin{array}{cc}1 & 0 \\1 & 1\end{array}\right)\,,~~~~
K^{2d}_{-\gamma^1} = \left(\begin{array}{cc}1 & 0 \\0 & (1+q^{-{1\over2}} X_{-\gamma^1})^{-1}\end{array}\right)\,,~~~~~~K^{4d}_{-\gamma^1}   =  \left(\begin{array}{cc}E_q(X_{-\gamma^1} )& 0 \\0 & E_q(X_{-\gamma^1})\end{array}\right)\,,~~~~~~
\end{split}
\end{align}
Note that the antiparticle in 2$d$ has opposite $C$ charge and degeneracy, $\omega_2(-\gamma^1, -1 ) =1$.  We will often combine $K^{2d}_{\gamma^1}$ with $K^{4d}_{\gamma^1}$ to form
\begin{align}
K_{\gamma^1}  \equiv K^{2d}_{\gamma^1}  K^{4d} _{\gamma^1} =   \left(\begin{array}{cc}E_q(X_{\gamma^1} )& 0 \\0 & E_q(qX_{\gamma^1})\end{array}\right)\,.
\end{align}
  
Let us apply our IR formula \eqref{IRsurfaceform} for the $2d$-$4d$ Schur index in this chamber,
\begin{align}
\mathcal{I}_\mathbb{S}(q)&=(q)_\infty^{2}  \text{Tr} \left[ {S}_{12;0} {K}_{\gamma^1}{K}_{\gamma^2}{S}_{\gamma_{21};0}    {K}_{-\gamma^1}{K}_{-\gamma^2}  \right]\notag\\
&= (q)_\infty^{2}  
\text{Tr} \left[ 
E_q( X_{\gamma^1})E_q(X_{\gamma^2} )
E_q( X_{-\gamma^1})  E_q(X_{-\gamma^2})\right.\notag\\
&\left.
-E_q(q X_{\gamma^1})E_q(X_{\gamma^2} )
E_q( X_{-\gamma^1})  E_q(X_{-\gamma^2})
+E_q(q X_{\gamma^1})E_q(X_{\gamma^2} )
E_q(q^{-1} X_{-\gamma^1})  E_q(X_{-\gamma^2})
\right]\,\notag\\
&=(q)_\infty^2 \sum_{\ell_1,\ell_2=0}^\infty {q^{\ell_1+\ell_2+\ell_1\ell_2 }\over [(q)_{\ell_1}(q)_{\ell_2}]^2}(2-q^{\ell_1}) 
=1+q+q^2+q^3+2q^4+2q^5+3q^6+\cdots\,.
\,.
\end{align}
The final answer is to insert $2-q^{\ell_1}$ into the double-sum formula of the Schur index without the defect, which is \cite{Cordova:2015nma}
\begin{align}
\mathcal{I}_{A_2}(q) = (q)_\infty^2 \sum_{\ell_1,\ell_2=0}^\infty {q^{\ell_1+\ell_2+\ell_1\ell_2 }\over [(q)_{\ell_1}(q)_{\ell_2}]^2}\,.
\end{align}
  We have checked that the surface defect index $\mathcal{I}_\mathbb{S}(q)$ agrees with the character of the $h=-1/5$ primary $\Phi_{1,2}$ in the $(2,5)$ minimal model to $\mathcal{O}(q^{100})$,
\begin{align}
\mathcal{I}_\mathbb{S}(q) =  \chi^{(2,5)}_{(1,2)} (q) \,.
\end{align}
Here the character $\chi^{(p,p')}_{(s,r)}(q)$ of the $\Phi_{s,r}$  primary  ($1\le s\le p-1$, $1\le r \le p'-1$) in the $(p,p')$ Virasoro minimal model is given by (e.g. see \cite{philippe1997conformal})
\begin{align}\label{minimalcharacter}
\begin{split}
\chi_{(s,r)}^{(p,p')}(q)  = q^{- {(r p-s p')^2 - (p-p')^2\over 4pp'} + {1\over 24}(1-{6(p-p')^2\over pp'}) } \left(\, K^{(p,p')}_{s,r} (q)   - K^{(p,p')}_{-s,r}(q)\, \right)
\end{split}
\end{align}
where
\begin{align}
\begin{split}
K^{(p,p')}_{s,r} ( q)  ={q^{-{1\over24}} \over (q)_\infty} \sum_{n\in \mathbb{Z} }  q^{  (2pp' n+ pr -p' s)^2 \over 4pp' }\,.
\end{split}
\end{align}
We have normalized the character to start from 1.  Note the following identification between the primaries in the minimal models, $\Phi_{s,r} = \Phi_{ p -s, p'-r}$.  In particular, the $(2,2n+3)$ Virasoro minimal model has  $n+1$ primaries (including the vacuum), $\Phi_{1,r}$ with $r=1,2,\cdots,n+1$.

We will reproduce this answer by Higgsing the $D_5$ Argyres-Douglas theory when we discuss the general case of $A_{2n}$.

\subsubsection{A Relation Between  Surfaces and Lines}

In \cite{Cecotti:2010fi, Cordova:2016uwk}, a surprising relation between the line defect index and the chiral algebra characters was found in many examples.  It was observed that even though the line defects do not preserve the chiral algebra cohomology, their Schur indices can often be written as linear combinations of characters of the associated chiral algebra. The simplest example is the $A_2$ Argyres-Douglas theory, where the line defect index is found to be equal to 
\begin{align}\label{linerelation}
\mathcal{I}_L(q)  = q^{-{1\over 2}} \chi_{(1,1)}^{(2,5)}(q) -q^{-{1\over 2}} \chi_{(1,2)}^{(2,5)}(q) \,,
\end{align}
where $\chi_{(1,1)}^{(2,5)}(q)$ and $\chi_{(1,2)}^{(2,5)}(q)$ are respectively the characters for the vacuum and weight $h=-1/5$ primary in the $(2,5)$ Virasoro minimal models.

This observation can be explained by a relation between the surface defects, whose indices are the chiral algebra characters, and the line defects as elaborated in section 4 of \cite{Cordova:2017ohl}.  More explicitly, we can cut open a surface defect by inserting an identity interface.  We can then  unwrap the surface defect into a sum of line defects.  A powerful advantage of the infrared formula \eqref{IRsurfaceform} for the surface defect index is that it respects this unwrapping process.  In this section we illustrate this procedure in the $A_2$ Argyres-Douglas theory and derive the relation \eqref{linerelation}.

Let us repeat the IR calculation in an alternative way such that the connection to the line defects will become clear.  Let us apply the refined wall-crossing formula $S_{12;\gamma} K_{ \gamma'}  =  K_{\gamma'} S_{12;\gamma+\gamma'} S_{12;\gamma}$ derived in section 3.3.1 of \cite{Cordova:2017ohl} to rewrite  the quantum spectrum generator as,
\begin{align}
\mathcal{S}^{2d-4d} _{\vartheta ,\vartheta+\pi }  = S_{12;0} K_{\gamma^1} K_{\gamma^2} 
= K_{\gamma^1} S_{12;\gamma^1}  S_{12;0} K_{\gamma^2} 
= K_{\gamma^1} S_{12;\gamma^1}  K_{\gamma^2}S_{12;0} \,.
\end{align}
Next, we apply the same basic wall-crossing formula to group the 4$d$ $K$-factors together,
\begin{align}
\mathcal{S}^{2d-4d} _{\vartheta ,\vartheta+\pi }  =
K_{\gamma^1} K_{\gamma^2} S_{12;\gamma^1+\gamma^2} S_{12;\gamma^1} S_{12;0}\,.
\end{align}
 Now we can rewrite our infrared formula for the surface defect index as,
\begin{align}
\mathcal{I}_\mathbb{S}(q)&
=(q)_\infty^{2}  \text{Tr} \left[ 
\Sigma^{2d}  {K}_{-\gamma^1}{K}_{-\gamma^2} {K}_{\gamma^1} {K}_{\gamma^2}   \right]\,,
\end{align}
where
\begin{align}
&\Sigma^{2d}\equiv  S_{12;\gamma^1+\gamma^2}S_{12;\gamma^1}{S}_{12;0}{S}_{21;0}   =
\left(\begin{array}{cc}
-q^{1\over2}  F(L_4) &
-1-q^{1\over2}  F(L_4) \notag\\
1 & 1\end{array}\right)\,,\\
&  {K}_{-\gamma^1}{K}_{-\gamma^2} {K}_{\gamma^1} {K}_{\gamma^2}  =
\left(\begin{array}{cc}
\mathcal{O}(X_{\gamma^1}, X_{\gamma^2})
&0 \\
0 &\mathcal{O}(qX_{\gamma^1}, X_{\gamma^2})\end{array}\right)\,.
\end{align}
Here $\Sigma^{2d}$ is the product of the $S$-factors coming from 2$d$ solitons and $\mathcal{O}(X_{\gamma^1} , X_{\gamma^2})$ is the 4$d$ quantum spectrum generator over the full central charge plane,
\begin{align}
\mathcal{O}(X_{\gamma^1} , X_{\gamma^2})  \equiv\mathcal{S}_{\vartheta,\vartheta+\pi}(q)\mathcal{S}_{\vartheta+\pi,\vartheta+2\pi}(q)
=E_q( X_{-\gamma^1})  E_q(X_{-\gamma^2} )  E_q(X_{\gamma^1} ) E_q(X_{\gamma^2})\,.
\end{align}
Above we show the dependence on $X_{\gamma^i}$ explicitly because sometimes they might be shifted by powers of $q$ due to the 2$d$ BPS particles.  Finally, $F(L_4)=X_{\gamma^1}+X_{\gamma^1+\gamma^2}$  is the generating function for the bulk line defect $L_4$ in the $A_2$ Argyres-Douglas theory \cite{Gaiotto:2010be,Cordova:2013bza}.

 From the above expression, we see that the surface defect index naturally decomposes into two terms, each of which can be interpreted as a line defect index,
\begin{align}\label{surfaceline}
\mathcal{I}_\mathbb{S} (q)  &=
  (q)_\infty^2 \text{Tr}\left[  \left(1-q^{1\over2} F(L_4)\right) \mathcal{O}(X_{\gamma^1} ,X_{\gamma^2})\right]
  \notag\\
  &= \mathcal{I}(q) -q^{1\over 2} \mathcal{I}_L(q) \,,
\end{align}
where in the last line we used the infrared formula for the line defect Schur indices \cite{Cordova:2016uwk}.  Finally by noting that $\mathcal{I}_\mathbb{S}(q)  = \chi^{(2,5)}_{(1,2)}(q)$ and $\mathcal{I}(q) = \chi^{(2,5)}_{(1,1)}(q)$, the above relation exactly reproduces the observation \eqref{linerelation} in \cite{Cordova:2016uwk} that the line defect Schur indices are linear combinations of the chiral algebra characters.

\subsubsection{Indices from BPS States: The $A_{2n}$ Theory}

The more general $A_{2n}$ Argyres-Douglas theory has the $A_{2n}$ Dynkin diagram as its BPS quiver.   It has a complex $n$-dimensional Coulomb branch and no flavor symmetry.  The associated chiral algebra is the $(2,2n+3)$ Virasoro minimal model \cite{Cordova:2015nma}.  We will see that the canonical surface defect index agrees with the character for the primary $\Phi_{1,2}$ in all cases. The other characters in the minimal model will arise from other surface defect indices. 

As studied in \cite{Gaiotto:2011tf}, there is one chamber on the Coulomb branch where the $2d$-$4d$ BPS  spectrum is (in increasing phase order)
\begin{align}
\gamma_{12}\,, \gamma^1\,,\gamma^3\,,\cdots\,, \gamma^{2n-1}\,, \gamma^2\,,\gamma^4\,,\cdots\,,\gamma^{2n}\,.
\end{align}
Here $\gamma^i$'s are a basis for the charge lattice and have the following Dirac pairings,
\begin{align}\label{diracp}
\langle\gamma^i,\gamma^j\rangle=
(-1)^{i+1} ( \delta_{j,i+1}+  \delta_{j,i-1})\,.
\end{align}
There is a $C$ charge neutral 2$d$ soliton interpolating between the two vacua and a unit $C$ charge 2$d$ particle with 4$d$ gauge charge $\gamma^1$ living in the second vacuum $\omega_2( \gamma^1,1)=-1$. 

Applying our IR formula for the canonical surface defect index, we obtain
\begin{align}
&\mathcal{I}_\mathbb{S}(q)=(q)_\infty^{2n}  \text{Tr} \left[ \mathcal{S}_{\gamma_{12}}\prod_{I:\text{odd}}  \mathcal{K}_{\gamma^I}\prod_{J:\text{even}} \mathcal{K}_{\gamma^J} \mathcal{S}_{\gamma_{21}}  \prod_{I:\text{odd}}  \mathcal{K}_{-\gamma^I}\prod_{J:\text{even}}\mathcal{K}_{-\gamma^J}  \right]\notag\\
= &(q)_\infty^{2n}  
\text{Tr} \left[ 
E_q(X_{\gamma^1})\prod_{I:\text{odd}-\{1\}} E_q(X_{\gamma^I})  \prod_{J:\text{even}} E_q(X_{\gamma^J} )
E_q( X_{-\gamma^1})\prod_{I:\text{odd}-\{1\}} E_q(X_{-\gamma^I})\prod_{J:\text{even}} E_q(X_{-\gamma^J} )\right.\notag\\
&\left.
-E_q(q X_{\gamma^1})\prod_{I:\text{odd}-\{1\}} E_q(X_{\gamma^I})\prod_{J:\text{even}} E_q(X_{\gamma^J} )
E_q(X_{-\gamma^1})\prod_{I:\text{odd}-\{1\}}  E_q(X_{-\gamma^I})\prod_{J:\text{even}} E_q(X_{-\gamma^J})\right.\notag\\
&\left.+E_q(q X_{\gamma^1})\prod_{I:\text{odd}-\{1\}} E_q(X_{\gamma^I})\prod_{J:\text{even}} E_q(X_{\gamma^J} )
E_q(q^{-1} X_{-\gamma^1})\prod_{I:\text{odd}-\{1\}} E_q(X_{-\gamma^I})\prod_{J:\text{even}} E_q(X_{-\gamma^J})
\right]\,\notag\\
&= (q)_\infty^{2n}\sum_{\ell_1,\cdots,\ell_{2n}=0 }^\infty
{ q^{  \sum_{i=1}^{2n} \ell_i  \,+ \sum_{i=1}^{2n-1} \ell_i \ell_{i+1}} \over\prod_{i=1}^{2n} [(q)_{\ell_i}]^2 }(2-q^{\ell_1})\,.
\end{align}
The final answer is to insert $2-q^{\ell_1}$ into the multiple-sum formula of the Schur index without the defect $\mathcal{I}_{A_{2n}}(q)$, which is \cite{Cordova:2015nma}
\begin{align}
\mathcal{I}_{A_{2n}}(q)= (q)_\infty^{2n}\sum_{\ell_1,\cdots,\ell_{2n}=0 }^\infty{ q^{  \sum_{i=1}^{2n} \ell_i  \,+ \sum_{i=1}^{2n-1} \ell_i \ell_{i+1}} \over\prod_{i=1}^{2n} [(q)_{\ell_i}]^2 }
\,.
\end{align}
 We have checked that the answer agrees with the character of the $\Phi_{1,2}$ primary for the $(2,2n+3)$ Virasoro minimal model to high orders in $q$,
\begin{align}
\mathcal{I}_\mathbb{S}(q) = \chi^{(2,2n+3)}_{(1,2)}(q)\,.
\end{align}

\subsubsection{Indices from Higgsing}
\label{sec:a2nhiggsing}

The $A_{2n}$ Argyres-Douglas theory can be realized by Higgsing the $D_{2n+3}$ theory, and the vortex surface defects in the former theory are realized by turning on position-dependent Higgs field in the latter.   Consequently, the $A_{2n}$ vortex surface defect indices can be obtained by taking the residues (in the flavor fugacity) of the $D_{2n+3}$ Schur index \cite{Gaiotto:2012xa}.  In addition to the canonical surface defect (which has unit vortex number), we will compute the indices for surface defects of any vortex number, which in turn reproduce all and only the characters of the associated $(2,2n+3)$ Virasoro minimal model. 

The Schur index of the $D_{2n+3}$ Argyres-Douglas theory is the vacuum character of 
$\widehat{SU(2)}_{-\frac{4(n+1)}{2n+3}}$ \cite{Buican:2015ina},
\begin{equation}
\mathcal{I}_{D_{2n+3}}(q,x)=\frac{1}{\prod_{k=1}^\infty (1-q^k)(1-x^2 q^k)(1-x^{-2} q^k)} \sum_{k=0}^\infty (-1)^k \frac{x^{2k+1}-x^{-2k-1}}{x-x^{-1}}q^{\frac{k(k+1)}{2}(2n+3)}\,.
\end{equation}
Notice the structure of an alternating sum of characters of Weyl modules of spin $0$, $1$, $2$, etc. 

The $SU(2)$ WZW currents are associated to the moment map operators for the $SU(2)$ flavor symmetry of the theory. We Higgs the theory by giving nilpotent vevs to these moment maps. 
This corresponds to the poles in the $SU(2)$ flavor fugacity $x$ at $x^{\pm1}=q^{(s+1)/2}$ for $s$ a non-negative integer in the index and to the usual qDS reduction 
in the chiral algebra. 

As a warm up, let us start by considering the  residue at the lowest order pole $x=q^{1/2}$. This corresponds to turning on a constant Higgs vev of the $D_{2n+3}$ theory.  In the IR the theory flows to the $A_{2n}$ theory without vortices.  The Schur index of the IR theory is captured by the residue at $x=q^{1/2}$  (normalizing by  $2(q)_\infty^2$),
\begin{align}
\mathcal{I}_{A_{2n}}(q)&=2(q)_\infty^2\text{Res}_{x\to q^{1/2}}{1\over x}\mathcal{I}_{D_{2n+3}}(q,y) \notag\\
&= \frac{1}{\prod_{k=2}^\infty (1-q^k)} \sum_{k=0}^\infty (-1)^k \frac{q^{k+\frac12}-q^{-k-\frac12}}{q^{\frac12}-q^{-\frac12}}q^{\frac{k(k+1)}{2}(2n+3)}\notag\\
&=\frac{1}{(q)_\infty} \sum_{\ell=-\infty}^\infty (q^{2\ell^2(2n+3)+(2n+1)\ell}-q^{2\ell^2(2n+3)+(2n+5)\ell+1} )
=\chi^{(2,2n+3)}_{(1,1)}(q)\,.
\end{align}
The answer  is precisely the vacuum character for the $(2,2n+3)$ Virasoro minimal model, which is indeed the Schur index of the $A_{2n}$ Argyres-Douglas theory \cite{Cordova:2015nma}. 

It is interesting how the sum over characters of Weyl modules of spin $k$ for the Kac-Moody algebra maps to a sum over characters of Virasoro modules with a single null vector at 
level $2k+1$, which is then reorganized to the alternating sum over character of Virasoro Verma modules which defines the $(2,2n+3)$ Virasoro vacuum character. 
It should be interesting to follow this correspondence at the level of the qDS reduction of the chiral algebra. 

More generally, the residue at $x = q^{(r+1)/2}$ corresponds to turning on a position-dependent Higgs vev. In the IR the system flows to the $A_{2n}$ Argyres-Douglas theory with a  surface defect, denoted by  $\mathbb{S}_r$, of vortex number $r$. The residue of the $D_{2n+3}$ index at $x = q^{(r+1)/2}$ then computes its surface defect index,\begin{align}
&\mathcal{I}_{\mathbb{S}_r}(q)
=2(-1)^{p_1+r} q^{p_2 -{r(r+1)\over2} } (q)_\infty^2 \text{Res}_{x\to q^{r+1\over2}}{1\over x}\mathcal{I}_{D_{2n+3}}(q,x)\notag\\
&=(-1)^{p_1+r} q^{p_2-{r(r+1)\over2}} \frac{(1-q^{r+1}) \cdots (1-q)}{(1-q^{-r})\cdots(1-q^{-1}) (q)_\infty} \sum_{k=0}^\infty (-1)^k \frac{q^{\frac{r+1}{2}(2k+1)}-q^{-\frac{r+1}{2}(2k+1)}}{q^{\frac{r+1}{2}}-q^{-\frac{r+1}{2}}}q^{\frac{k(k+1)}{2}(2n+3)}\notag\\
&=\frac{(-1)^{p_1}q^{p_2}}{(q)_\infty}\sum_{\ell=-\infty}^\infty (q^{2\ell^2(2n+3)+(2n-2r+1)\ell}-q^{2\ell^2(2n+3)+(2n+2r+5)\ell+ r+1})\,,
\end{align}
where the overall factor $2(-1)^{p_1+r} q^{p_2 -{r(r+1)\over2} } $ is inserted to normalized the index to start from 1.  
The subscript $r$ labels the vortex number of the surface defect.  
Here $p_1=\lfloor {r\over 2n+3}\rfloor$ and $p_2= {1\over2}\lfloor {r\over 2n+3}\rfloor \left[
(2n+3)\lfloor {r\over 2n+3}\rfloor- 2n-1+2 \bar r\right]$\,.  We have defined $\bar r\equiv r\,\,\text{mod}\,\,(2n+3)$.   

  Remarkably, these vortex surface defect indices reproduce \textit{all and only} the $n+1$ characters of the $(2,2n+3)$ Virasoro minimal model,
  \begin{align}
  \mathcal{I}_{\mathbb{S}_r}(q) = \chi^{(2, 2n+3)}_{(1,  \bar r+1)}(q)\,,~~~~~\bar r\equiv r\,\,\text{mod}\,\,(2n+3)\,.
  \end{align}
 In particular, the surface defect index has a periodicity of $2n+3$ in the vortex number $r$, and a reflection symmetry $\bar r\leftrightarrow 2n+1-\bar r$,  which mirrors the reflection symmetry of the $(p,p')$ minimal model primaries $\Phi_{s,r} = \Phi_{p-s,p'-r}$.  It vanishes if $\bar r+1$ is a multiple of $2n+3$.  Note that the canonical surface defect discussed previously has unit vortex number $r=1$ and gives rise to the character for  $\Phi_{1,2}$.  The resulting behavior of the vortex surface defect indices for the $A_{2}$ theory is illustrated in Table \ref{table}.
 
 \begin{table}[h]
\centering
\begin{tabular}{|c|c|c|c|c|c|c|c|}
\hline
Vortex Number $r$ & 0  &1 &2&3&4 &5 &$\cdots$\\
\hline
&&&&&&&\\
$\mathcal{I}_{\mathbb{S}_r}(q)$&$\chi_{(1,1)}^{(2,5)}$& $\chi_{(1,2)}^{(2,5)}$&
$\chi_{(1,2)}^{(2,5)}$&$\chi_{(1,1)}^{(2,5)}$&~~~0~~~&$\chi_{(1,1)}^{(2,5)}$  
&$\cdots$\\
&&&&&&&\\
\hline
\end{tabular}
\caption{The vortex surface defect indices in the Schur limit for the $A_2$ Argyres-Douglas theory.  $\chi_{(1,1)}^{(2,5)}$ and $\chi_{(1,2)}^{(2,5)}$ denote the characters of the vacuum and the $h=-1/5$ primary, respectively.  The defect Schur index in this case is periodic in the vortex number $r$ with period 5.  It also enjoys a reflection symmetry $\bar r \leftrightarrow 3-\bar r$, where $\bar r \equiv r ~\text{mod}~5$.}
\label{table}
\end{table}
 
 In the special case $n=0$, the bulk theory is empty, and the defect Schur index is either $1$ or 0 depending on the vortex number $r$:
   \begin{align}\label{emptyschur}
  \mathcal{I}_{\mathbb{S}_r}(q) = 
  \begin{cases}
  &1, ~~~~~~~~r = 0,1~~\text{mod}~3\,,\\
    &0,~~~~~~~~ r = 2~~~~~\text{mod}~3\,.
  \end{cases}
  \end{align}

The previous discussion shows that the $2d$-$4d$ Schur index cannot completely distinguish between the vortex surface defects $\mathbb{S}_{r}$.  This then presents the question: are these vortex defects physically the same, or is the $2d$-$4d$ Schur index simply to coarse an observable?  We can probe this question by computing more refined invariants of the defects.
\begin{itemize}
\item We can look at the details of the qDS reduction to probe if the corresponding chiral algebra modules also have the periodicity and reflection symmetry manifested by the characters.
\item We can compute the full superconformal index of the surface defect if we are given the full superconformal index of the $D_{2n+3}$ theory. 
\end{itemize}
These two ideas are not unrelated.  Indeed, the Macdonald limit of the superconformal index receives contributions from the same set of operators as the Schur index \cite{Gadde:2011uv}.\footnote{Technically, this has only been demonstrated for the index without a surface defect.  It is natural to expect that it continues to hold in the presence of defects as well.}  Thus if the Macdonald indices are distinct, the chiral algebra modules are also distinct.  Motivated by this, in  appendix \ref{sec:emptyvortices} we calculate the full index in a very special example: the reduction of the $D_3$ theory to a trivial theory. That example already shows the lack of periodicity and reflection symmetry in the Macdonald defect index suggesting that the chiral algebra modules are indeed distinct.  It would be interesting to reproduce this directly from qDS reduction.

\subsection{$A_{2n+1}:$  $W^{(2)}_{n+1}$  Algebras}

The $4d$ $A_{2n+1}$ Argyres-Douglas superconformal theory has a complex $n$-dimensional Coulomb branch and an $U(1)$ flavor symmetry. The flavor symmetry is enhanced to $SU(2)$ in the cases of $n=0,1$.  In particular, the $A_1$ Argyres-Douglas theory is the free hypermultiplet discussed in detail in section \ref{sec:freehyper}. 

The chiral algebra of the $A_{2n+1}$ Argyres-Douglas theory is the $W^{(2)}_{n+1}$  algebra at level $k=-(n+1)^2/(n+2)$ \cite{beem,Creutzig:2017qyf}.  The $W^{(2)}_{n+1}$  algebra at level $k$ is  the quantum Hamiltonian reduction corresponding to a certain non-principal embedding of $SU(2)$ into $\widehat{SU(n+1)}_k$ \cite{Feigin:2004wb}. The series of $W^{(2)}_{n+1}$ algebras at level $k=-(n+1)^2/(n+2)$ for $n=0,1,2,\cdots$ is studied in \cite{Creutzig:2013pda,Creutzig:2017qyf}.

Let us discuss the first few cases of the $A_{2n+1}$ chiral algebras.

\subsubsection{$A_3:\widehat{SU(2)}_{-\frac43}$}
The $A_3$ chiral algebra is $\widehat{SU(2)}_{-\frac43}$, whose vacuum character has been matched with the Schur index of the $A_3$ Argyres-Douglas theory \cite{Buican:2015ina,Cordova:2015nma}.  Recall that the nonvanishing commutation relations in the $\widehat{SU(2)}_k$ Kac-Moody algebra are
\begin{align}
\begin{split}
&[J^+_n  , J^-_m ] = 2 J^3_{n+m}+ k n \delta_{n,-m}\,,\\
&[J^3 _n , J^\pm_m ]  =\pm J^\pm_{n+m}\,,\\
&[J^3_n,J^3_m]= {k\over 2} n \delta_{n,-m}\,.
\end{split}
\end{align}
The $\widehat{SU(2)}_k$ has a spectral flow symmetry labeled by $\eta\in \mathbb{R}$ that leaves the commutation relations invariant. The spectral flow by $\eta$ units acts on the currents as
\begin{align}
J^3_n \to J^{3'}_n=  J^3_n +{k\over2} \eta \delta_{n,0}\,,~~~~~~
J^\pm_n  \to J^{\pm ' }_n = J^\pm_{n\pm\eta}\,.
\end{align}
Let $\lambda$ be the Dynkin label  for the finite $SU(2)$ of a highest weight state and $h$ its  conformal weight determined by the Sugawara construction.\footnote{For an untwisted module, it is given by $h= {\lambda(\lambda+2)\over 4(k+2)}$.}  We can then produce a new highest weight module $(\lambda',h')$ from an existing one $(\lambda,h)$ by $\eta$ units of spectral flow, where
\begin{align}\label{spectralflow}
\lambda'  = \lambda + k \eta \,,~~~~~
h'=h+{\eta\over 2}\lambda +{k\over 4}\eta^2\,.
\end{align}
In particular, a spectral flow by a half integral amount shift $J^\pm_n$ to be half-integrally moded and leaves $J^3_n$ integrally moded.  Modules of the half-integrally spectral-flowed $\widehat{SU(2)}_k$ algebra will be called the twisted modules. Note that the twisted $\widehat{SU(2)}_k$ does \textit{not} contain the finite $SU(2)$ Lie algebra as its subalgebra since the $J^\pm_0$ zero modes are lifted under spectral flow.

The $\widehat{SU(2)}_{-\frac43}$  twisted module that  will be related to the canonical surface defect is the $\eta=-1/2$ spectral-flowed twisted module of an admissible untwisted module with $(\lambda = -\frac23, h =-\frac13)$.  The highest weight of this twisted module then has $\lambda= 0$ and $h=-{5\over 12}$ by \eqref{spectralflow}.  The  character of the untwisted $(\lambda = -\frac23, h =-\frac13)$ module  is (see, for example, \cite{philippe1997conformal})
\begin{align}
\chi_{(\lambda=-\frac 23,  h=-\frac13)}(q,w) = {w^{-2/3}\over 1-w^2} {1+\sum_{k=1}^\infty (-1)^k w^{2k}q^{\frac k2 (3k-1)} +w^{-2k} q^{\frac k2 (3k+1)}) \over \prod_{k=1}^\infty (1-q^k)(1-w^2 q^k)(1-w^{-2}q^k)}\,,
\end{align}
where $w$ is the $SU(2)$ flavor fugacity.  
This character has  appeared in the study of line defect indices of the $A_3$ Argyres-Douglas theory \cite{Cordova:2016uwk}. Using \eqref{spectralflow}, it follows that the character $\chi^{tw}_{(\lambda=0,  h=-\frac{5}{12})}(q,w)$ for the twisted module is  given by (up to an overall factor) replacing $w\to wq^{1\over4}$ in the untwisted character,
\begin{align}\label{twistedSU(2)}
\chi^{tw}_{(\lambda=0,  h=-\frac{5}{12})}(q,w) ={1+\sum_{k=1}^\infty (-1)^k (w^{2k}+w^{-2k} ) q^{3k^2\over2}\over \prod_{k=1}^\infty (1-q^{k})(1-w^2 q^{k-1/2})(1-w^{-2}q^{k-1/2})}\,,
\end{align}
where we have normalized the twisted character to start from 1. Notice the denominator representing all twisted and untwisted current algebra modes 
and the numerator encoding the structure of null vectors.

 We will see  that the canonical surface defect index in the $A_3$ Argyres-Douglas theory, computed both from the IR formula and the Higgsing procedure, equals the above twisted character.

\subsubsection{$A_5:$ Bershadsky-Polyakov Algebra $W^{(2)}_3$}

The Bershadsky-Polyakov algebra $W^{(2)}_3$ at level $k$  \cite{Bershadsky:1990bg,Polyakov:1989dm} contains a weight one bosonic generator $J(z)$, two weight $\frac32$ \textit{bosonic} generators $G^\pm(z)$, and one weight 2 bosonic generator $T(z)$. Note that this algebra almost has the same generators as the $\mathcal{N}=2$ Virasoro algebra except that in the case of $W^{(2)}_3,$ all the generators are bosonic.  The central charge is 
\begin{align}
c= 25- {24\over k+3} -6(k+3)\,.
\end{align}
The OPEs are
\begin{align}
\begin{split}
&J(z)J(0) \sim {2k+3\over 3z^2}\,,\\
&G^+(z) G^-(0) \sim {(k+1)(2k+3)\over z^3} + {3(k+1)\over z^2} J(0)\\
&~~~~~~~~~~~~~~~~~
+{1\over z}\left[3 :JJ:(0) -(k+3)T(0)+{3(k+1)\over 2}\partial J(0) \right]\,,
\end{split}
\end{align}
with the remaining OPEs determined by $U(1)$ charge conservation, $J(z)$ and $G^\pm(z)$ being  Virasoro primaries of weight 1 and 3/2, respectively, and $G^\pm(z)$ being current algebra primaries of charge $\pm1$.  Note that the Bershadsky-Polyakov algebra is nonlinear because of the $:JJ:$ term on the righthand side of the $G^+(z)G^-(0)$ OPE.

The $W^{(2)}_3$ algebra, similar to the $\mathcal{N}=2$ Virasoro algebra, has a spectral flow symmetry labeled by $\eta\in\mathbb{R}$, which acts as
\begin{align}
\begin{split}
&G^\pm_r \to G^{\pm'}_r=G^{\pm}_{r\pm\eta} \,,\\
&J_n \to J_n'  =  J_n +{2k+3\over3} \eta\delta_{n,0}\,,\\
&L_n \to L_n' = L_n +\eta J_n +  {2k+3\over 6} \eta^2 \delta_{n,0}\,.
\end{split}
\end{align} 
In particular, half-integer units of spectral flow will shift the moding of  $G^\pm_r$ from  half-integers to integers, and vice versa.  
If $G^\pm_r$ are integrally (half-integrally) moded, we will call the algebra and its modules twisted (untwisted).

The $A_5$ chiral algebra is the Bershadsky-Polyakov algebra $W^{(2)}_3$ at level $k=-9/4$, and hence $c=-23/2$ \cite{Creutzig:2017qyf, beem}.  The $U(1)$ current $J(z)$ descends from the $4d$ $U(1)$ flavor symmetry current of the $A_5$ Argyres-Douglas theory.    To provide evidence for the above claim, one should check the  $W^{(2)}_3$ vacuum character equals the $A_5$ Schur index without any defect. Using the \texttt{Weaver} \texttt{Mathematica} package of \cite{Whalen:2014fta}, we compute the first few terms of the vacuum character (turning off the $U(1)$ flavor fugacity)
\begin{align}\label{BPvac}
\chi_0^{W^{(2)}_3}(q,x=1) = 1+q+2q^{3\over2} +3q^2 +4q^{5\over2} +7q^3 + 8q^{7\over2} + 14q^4+\mathcal{O}(q^{9\over2})\,.
\end{align}
This nicely agrees with the $A_5$ Schur index derived in \cite{Buican:2015ina,Cordova:2015nma},\footnote{The character of $W_{n+1}^{(2)}$ at level $k=-(n+1)^2/(n+2)$ is recently shown to be equal to the Schur index of the $A_{2n+1}$ Argyres-Douglas theory for all $n$ in \cite{Creutzig:2017qyf}.}
\begin{align}
 \mathcal{I}_{A_5}(q,x=1)=\chi_0^{W^{(2)}_3}(q,x=1)  \,.
\end{align}
As we will see in this section, both the twisted and the untwisted characters of the $W^{(2)}_3$ algebra will appear as the surface defect indices of the $A_5$ Argyres-Douglas theory.

\subsubsection{Indices from BPS States}

The $2d$-$4d$ BPS spectrum for the $A_{2n+1}$ Argyres-Douglas theory with the canonical surface defect is analogous to the $A_{2n}$ Argyres-Douglast theory. The Dirac pairings between the $4d$ charges are again \eqref{diracp}. There is a flavor symmetry $U(1)$, which is enhanced to $SU(2)$ for the $A_3$ theory,  associated with the lattice vector $\gamma_f$ defined as
\begin{align}\label{A2n1flavor}
\gamma_f  =  \sum_{i=0}^n (-1)^i \gamma_{2i+1}\,.
\end{align}

We will be working  in a chamber where there is a $C$ charge neutral $2d$ soliton between the two vacua with degeneracy $\mu_{12}(0,0)=1$, as well as a unit $C$ charge  2$d$ BPS particle with 4$d$ charge $\gamma^1$ living in the second vacuum $\omega_2(\gamma^1 ,1)=-1$.  The $2d$-$4d$ BPS  spectrum is (in increasing phase order)
\begin{align}
\gamma_{12}\,, \gamma^1\,,\gamma^3\,,\cdots\,, \gamma^{2n-1}\,, \gamma^2\,,\gamma^4\,,\cdots\,,\gamma^{2n-2}\,,
\end{align}
where again $\gamma^1$ collectively stand for the 4$d$ particle and the 2$d$ particle living in the second vacuum.  
The IR calculation proceeds exactly as in the $A_{2n}$ case and we obtain
\begin{align}
& (q)_\infty^{2n} \sum_{\substack{\ell_1,\cdots,\ell_{2n+1},\\ k_1,\cdots ,k_{2n+1}=0} }^\infty
{(-1)^{\sum_{i=1}^{2n+1} (k_i+\ell_i) } q^{{1\over2}  \sum_{i=1}^{2n+1} (k_i+\ell_i ) + \sum_{j=1}^n  \ell_{2j}(\ell_{2j-1}+\ell_{2j+1}) } \over\prod_{i=1}^{2n+1} (q)_{k_i} (q)_{\ell_i}}\\
&\times\, [(-1)^{n+1}x]^{\ell_1-k_1}\,  
\left(1 +q^{\ell_1-k_1} -q^{\ell_1}\right)\,
\prod_{i=1}^n \delta_{k_{2i},\ell_{2i}} \, \prod_{j=1}^n  \delta_{(-1)^{j+1}k_1 + k_{2j+1} ,(-1)^{j+1}\ell_1 + \ell_{2j+1}  }\,,\notag
\end{align}
The $U(1)$ flavor fugacity  $x$ is normalized to be the trace of the flavor generator as
\begin{align}
\text{Tr}[X_{\gamma_f}] =
 (-1)^{n+1}x\,.
\end{align}
Finally we should shift the flavor fugacity by 
\begin{align}
x \to xq^{-\frac12}\,.
\end{align}
As discussed in section \ref{sec:monodromy}, this is modifying the surface defect in the above calculation by a monodromy twist.  In this way we obtain the canonical surface defect for the $A_{2n+1}$ Argyres-Douglas theory which resides in the twisted sector.  Hence
\begin{align}\label{Aoddcanonical}
&\mathcal{I}_\mathbb{S} (q,x) = (q)_\infty^{2n} \sum_{\substack{\ell_1,\cdots,\ell_{2n+1},\\ k_1,\cdots ,k_{2n+1}=0} }^\infty
{(-1)^{\sum_{i=1}^{2n+1} (k_i+\ell_i) } q^{{1\over2}  \sum_{i=1}^{2n+1} (k_i+\ell_i ) + \sum_{j=1}^n  \ell_{2j}(\ell_{2j-1}+\ell_{2j+1}) } \over\prod_{i=1}^{2n+1} (q)_{k_i} (q)_{\ell_i}}\\
&\times\, [(-1)^{n+1}x]^{\ell_1-k_1}\,  
\left(q^{\ell_1-k_1\over2} +q^{k_1-\ell_1\over2}-q^{\ell_1+k_1\over2} \right)\,
\prod_{i=1}^n \delta_{k_{2i},\ell_{2i}} \, \prod_{j=1}^n  \delta_{(-1)^{j+1}k_1 + k_{2j+1} ,(-1)^{j+1}\ell_1 + \ell_{2j+1}  }\,,\notag
\end{align}
The final answer is to insert $q^{\ell_1-k_1\over2} +q^{k_1-\ell_1\over2}-q^{\ell_1+k_1\over2}$ into the multiple-sum formula of the $A_{2n+1}$ Schur index $\mathcal{I}_{A_{2n+1}}(q,x)$ without the defect in \cite{Cordova:2015nma}.

For the $A_{3}$ Argyres-Douglas theory, the $U(1)$ flavor symmetry is enhanced to $SU(2)$, and we will  replace the $U(1)$ flavor fugacity $x$ by $w^2$ so that the character of the $n$-dimensional representation of $SU(2)$ is $\chi_\mathbf{n} (w)  =  ( w^{n} -w^{-n})/( w-w^{-1} )$. The first few terms of the canonical surface defect index are
\begin{align}
&\mathcal{I}_\mathbb{S}(q,w) = 1+ (w^2+w^{-2} ) q^{1\over2}   +  (w^4 +2+w^{-4} ) q 
+(w^6 + 2w^2  +2w^{-2} +w^{-6} )q^{3\over2} \notag\\
& +  (w^8 + 2w^4 +4+ 2w^{-4} +w^{-8})q^2 
+(w^{10}  + 2w^6 + 5w^2 + 5w^{-2} + 2w^{-6} +w^{-10} ) q^{5\over2}  \notag\\
&
+(w^{12} + 2w^8  +  5w^4  +8+ 5w^{-4}  + 2w^{-8}  + w^{-12} )q^3\notag\\
&
+(w^{14}  + 2w^{10} + 5w^6 + 10w^2  +10w^{-2} +5w^{-6} +2w^{-10} +w^{-14})q^{7\over2}
+\mathcal{O}(q^4)\,.
\end{align}
 We have checked that the above surface defect index agrees with the twisted character of $\widehat{SU(2)}_{-\frac43}$ with $\lambda=0$ \eqref{twistedSU(2)} to $\mathcal{O}(q^{9})$,
 \begin{align}\label{A3twist}
 \mathcal{I}_\mathbb{S}(q,w)  = \chi_{(\lambda=0,h=-\frac{5}{12})}^{tw}(q,w)\,.
 \end{align}
   Note that the coefficients of the index can \textit{not} be written as non-negative sums of $SU(2)$ characters because the $SU(2)$ flavor symmetry is broken by the canonical surface defect.  Correspondingly, the twisted $\widehat{SU(2)}_{-\frac43}$ does \textit{not} contain the finite $SU(2)$ Lie algebra as its subalgebra because the zero modes $J^\pm_0$ are lifted under spectral flow.

Moving on to the $A_5$ case, the first few terms of the canonical surface defect index are
\begin{align}
\mathcal{I}_\mathbb{S} ( q,x)&  =1 +  (x+2+x^{-1})q +  (x^2+2x +4+2x^{-1}+x^{-2})q^2\notag\\
&  + (x^3 +2x^2+5x +  8 +5x^{-1}  + 2x^{-2} +x^{-3})q^3 \\
&
+(x^4 +2x^3 + 5x^2+ 10x + 15+10x^{-1} +5x^{-2} +2x^{-3} +x^{-4})q^4\notag\\
&+
(x^5+ 2x^4+ 5x^3 + 10x^2+19x+26+19x^{-1} +10x^{-2} +5x^{-3} +2x^{-4}  +x^{-5})q^5
+\mathcal{O}(q^6)\notag\,.
\end{align}
which agrees with the twisted character with $L_0=-\frac{5}{16}$, $J_0=0$, $G^\pm_0=0$ of the Bershadsky-Polyakov algebra $W^{(2)}_3$ at level $k=-\frac94$.  We have checked the above claim using the \texttt{Weaver} \texttt{Mathematica} package of \cite{Whalen:2014fta} to $\mathcal{O}(q^5)$ with the flavor fugacity $x$ off,
\begin{align}\label{BPtwist}
\mathcal{I}_\mathbb{S}(q,x=1) =  \chi^{tw}_{L_0 = -\frac{5}{16},J_0=0}(q,x=1)\,.
\end{align}

To sum up, we computed the canonical surface defect indices  \eqref{Aoddcanonical} of the $A_{2n+1}$ Argyres-Douglas theories from the $2d$-$4d$ BPS states. We checked that the answers nicely reproduce the twisted characters of the associated chiral algebra $W^{(2)}_{n+1}$ for small $n$.  In the following we will compute the more general surface defect indices in these theories from the Higgsing procedure and match with the other characters of the associated chiral algebra.

\subsubsection{Indices from Higgsing}

The $A_{2n+1}$ Argyres-Douglas theory can be obtained by Higgsing the $D_{2n+4}$ theory.  At the level of the index,  the $A_{2n+1}$ vortex surface defect indices of all vortex numbers can be computed from the residues of the $D_{2n+4}$ Schur index without the defect.  
The Schur index of the $D_{2n+4}$ theory is \cite{Buican:2015ina}
\begin{align}
&\mathcal{I}_{D_{2n+4}} (q,x_1,x_2) \notag\\
&= { \left( \mathcal{I}^{SU(2)}_{\text{vect}}(x_2)\right)^{-\frac12}\over (q)_\infty }
\sum_{k=0}^\infty \left(
{1\over x_2-x_2^{-1}}\sum_{p=\pm1} \left[
{q^{(  2k^2 +2k+1 )(n+2) \over 2} x_2^{2k+2} x_1^p \over 1-q^{(n+2)(k+\frac12)}x_2 x_1^p }
-{q^{( 2k^2+4k+1)(n+2) \over 2}  x_2^{-2k-2} x_1^p \over 1-q^{(n+2)(k+\frac12)}x_2^{-1} x_1^p }
\right]\right.\notag\\
&\left.
+q^{(n+2)k(k+1)}\chi_{2k+1}(x_2)
\right)\,,
\end{align}
where
\begin{align}
\left( \mathcal{I}^{SU(2)}_{\text{vect}}(x_2)\right)^{-\frac12}
={1\over (q)_\infty }  \prod_{m=1}^\infty {1\over (1-q^m x_2^2 )(1-q^m x_2^{-2})}\,,
\end{align}
and $\chi_n(x_2)$ is the $SU(2)$ character for the $n$-dimensional representation.  
The $D_{2n+4}$ Argyres-Douglas theory has $SU(2)\times U(1)$ flavor symmetry.  Here $x_1$ is the $U(1)$ fugacity and $x_2$ is the $SU(2)$ fugacity.  The $D_{2n+4}$ index has poles in the $SU(2)$ fugacity at $x_2=q^{r+1\over 2}$ ($r=0,1,2,\cdots)$, whose residues give the vortex defect indices of the $A_{2n+1}$ theory.  
It follows that the $2d$-$4d$ Schur index  of a surface defect $\mathbb{S}_r$ of vortex number $r$ in the $A_{2n+1}$ theory is,
\begin{align}\label{higgsDeven}
&\mathcal{I}_{\mathbb{S}_r} (q,x) 
= 2(q)^2_\infty (-1)^{p_1+r} q^{p_2-{r(r+1)\over2}} \text{Res}_{x_2=q^{r+1\over 2} }   {1\over x_2}   \mathcal{I}_{D_{2n+4}} (q,x,x_2)\notag\\
&={(-1)^{p_1}\over (q)_\infty^2}  q^{p_2}(1-q^{r+1})
\sum_{k=0}^\infty \left(
{1\over q^{r+1\over2} -q^{-{r+1\over2} }}\sum_{p=\pm1} \left[
{q^{{ (2k^2+4k+1)(n+2)\over 2 }  +(k+1)(r+1)}  x^p \over 1-q^{(n+2)(k+\frac12) + {r+1\over2} } x^p }
\right.\right.\notag\\
&\left.\left.
-{q^{{(2k^2+4k+1)(n+2)\over 2} - (k+1)(r+1)}   x^p \over 1-q^{(n+2)(k+\frac12)- {r+1\over2} } x^p }
\right]
+q^{(n+2)k(k+1)}  {q^{(k+\frac12)(r+1)}  -q^{-(k+\frac12)(r+1) } \over q^{r+1\over2 }- q^{-\frac{r+1}{2}}}
\right)\,
\end{align}
where $p_1= \left\lfloor {r\over n+2} \right\rfloor$ and $p_2 =  (n+2) \left( \left\lfloor {r\over 2n+4}\right \rfloor \right)^2 -\left(n+1-\bar r \right)\left\lfloor {r\over 2n+4}\right\rfloor$.   Here $\bar r \equiv  r\, \text{mod}\,(2n+4)$.  
The factor $2(q)^2_\infty (-1)^{p_1+r} q^{p_2-{r(r+1)\over2}} $ is  inserted to normalize the index to start from 1.  Similar to the case in the $A_{2n}$ theory discussed in section \ref{sec:a2nhiggsing},  the vortex surface defect index $\mathcal{I}_{\mathbb{S}_r} (q,x_1)$ is periodic in $r$ with periodicity $2n+4$, i.e. $\mathcal{I}_{\mathbb{S}_{r+2n+4}} (q,x_1) = \mathcal{I}_{\mathbb{S}_r} (q,x_1)$.  It also enjoys the reflection symmetry $\bar r \leftrightarrow 2n+2-\bar r$, i.e. $\mathcal{I}_{\mathbb{S}_{\bar r}} (q,x_1) = \mathcal{I}_{\mathbb{S}_{2n+2-\bar r}} (q,x_1)$. Furthermore, it vanishes for $\bar r= n+1$ and $\bar r=2n+3$.

We expect that the phenomena we saw for the free hypermultiplet will happen generically and that the periodicity and zeroes of the Schur indices of vortex defects in $A_{2n+1}$ theories do not actually not hold at the level of chiral algebra modules, similarly to the case of $A_{2n}$ theories.  Indeed this is born out by the Macdonald index calculations of appendix \ref{sec:fullhyper}.

For the $A_3$ theory, we have checked to high orders of the $q$-expansion that (recall that in the $A_3$ theory it is more natural to use the $SU(2)$ fugacity $w=\sqrt{x}$) 
\begin{align}
\mathcal{I}_{\mathbb{S}_r} (q,w) =  
\begin{cases}
\chi_0^{\widehat{SU(2)}_{-4/3}}(q,w)\,~~~~~~~~~\text{if}~~~ r= 0,4\,\text{mod}\,6\,,\\
\chi_{(\lambda=0 , h=-\frac{5}{12})}^{tw}(q,w)\,~~~~~~\text{if}~~~ r= 1,3\,\text{mod}\,6\,,\\
0~~~~~~~~~~~~~~~~~~~~~~~\,~~~~\text{if}~~~r=2,5\,\text{mod}\,6\,,
\end{cases}
\end{align}
where $\chi_0^{\widehat{SU(2)}_{-4/3}}(q,w)$ and $\chi_{(\lambda=0 , h=-\frac{5}{12})}(q,w)$ are the vacuum character (see, for example, \cite{Buican:2015ina,Cordova:2015nma} for the explicit expression) and the twisted character \eqref{twistedSU(2)} of $\widehat{SU(2)}_{-\frac43}$, respectively.  In particular, the surface defect with unit vortex number $r=1$ is the canonical surface defect and the above results reproduce the answer \eqref{A3twist} obtained from the IR formula \eqref{IRsurfaceform}.

For the $A_5$ case,  we have checked to high orders of the $q$-expansion that
\begin{align}
\mathcal{I}_{\mathbb{S}_r} (q,x) =  
\begin{cases}
\chi_0^{W^{(2)}_3}(q,x)\,~~~~~~~~~~~~\,~~~\text{if}~~~ r= 0,6\,\text{mod}\,8\,,\\
 \chi^{tw}_{L_0 = -\frac{5}{16},J_0=0}(q,x)~~~~~~~\text{if}~~~ r= 2,4\,\text{mod}\,8\,,\\
 \chi_{L_0 = -\frac{1}{2},J_0=0}(q,x)~~~~~~~~\text{if}~~~r=1,5\,\text{mod}\,8\,,\\
 0~~~~~~~~~~~~~~~~~~~~~~\,~~~~~~\text{if}~~~r=3,7\,\text{mod}\,8\,,
\end{cases}
\end{align}
where $\chi_0^{W^{(2)}_3}(q,x)$ and $ \chi^{tw}_{L_0 = -\frac{5}{16},J_0=0}(q,x)$ are the aforementioned  vacuum \eqref{BPvac} and  twisted characters \eqref{BPtwist} of $W^{(2)}_3$ at $k=-\frac94$, respectively.  The other index $\chi_{L_0 = -\frac{1}{2},J_0=0}(q,x)$ is a another untwisted character of $W^{(2)}_3$ with $L_0=-\frac12$ and $J_0=0$.  We computed this untwisted character again using the \texttt{Weaver Mathematica} package \cite{Whalen:2014fta} to $\mathcal{O}(q^{4})$ (with the $U(1)$ fugacity off) and verify it agrees with $\mathcal{I}_{\mathbb{S}_r} (q,x=1)$ for $r=1,5\,\text{mod}\,8$.

 \section*{Acknowledgements} 
We thank Tomoyuki Arakawa, Chris Beem, Thomas Creutzig,  Andy Neitzke, Wolfger Peelaers, Leonardo Rastelli, and Fei Yan for interesting discussions.  We thank Wolfger Peelaers for comments on a draft.  The work of CC is supported by a Martin and Helen Chooljian membership at the Institute for Advanced Study and DOE grant DE-SC0009988. The research of DG was supported by the Perimeter Institute for Theoretical Physics.  Research at Perimeter Institute is supported by the Government of Canada through Industry Canada and by the Province of Ontario through the Ministry of Economic Development \& Innovation.  SHS is supported by the National Science Foundation grant PHY-1606531.

\appendix

\section{The Full Superconformal Index for Vortex Defects in an Empty $4d$ Theory}
\label{sec:emptyvortices}

In this appendix we compute the full, three-variable superconformal index for the surface defects in an empty $4d$ theory.  Even though the $4d$ theory is empty, we can introduce surface defects  with interacting degrees of freedom, and hence make the $2d$-$4d$ system nontrivial  
We will focus on vortex defects $\mathbb{S}_r$ arising from turning on a position-dependent Higgs field in the UV $D_3\cong A_3$ Argyres-Douglas theory, and then flowing to the IR  empty bulk theory with surface defects.  The superconformal indices for these vortex defects can be obtained as the residues of the full superconformal index of the $D_3$ theory without the defect \cite{Gaiotto:2012xa} (see section \ref{sec:higgs} for a more detailed discussion).

The full superconformal index of the $D_3$ theory is \cite{Maruyoshi:2016aim,Agarwal:2016pjo}\footnote{There are two equivalent expressions for the index depending on whether we think of it as the $D_3$ or $A_3$ theory. 
The pole structure is more manifest for the purpose of Higgsing if we view it as a $D_3$ theory.}
\begin{align}\label{D3full}
&\mathcal{I}^{D_3}  (p,q,t,a) 
={\kappa\over2} {\Gamma\left(( {pq/t})^{4\over3}  \right) \Gamma\left(  (pq/t)^{1\over3} \right) \over\Gamma\left(( pq/t)^{2\over3} \right)}
\oint {dz\over 2\pi i z} {\Gamma\left(z^{\pm 2} (pq)^{1\over3} t^{-\frac13} \right) \over \Gamma(z^{\pm2})}\notag\\
&\times
 \Gamma\left(z^\pm a^\pm (pq)^{\frac 13} t^{1\over6} \right) \,
\Gamma\left(z^\pm (pq)^{\frac 13} t^{1\over6} \right)\,
\Gamma\left(z^\pm  (pq)^{-\frac 13} t^{5\over6} \right)
\,,
\end{align}
where
\begin{align}
&\Gamma(z) \equiv \prod_{m,n=0}^\infty {1-z^{-1}p^{m+1}q^{n+1} \over 1-zp^m q^n }\,,
\end{align}
is the elliptic gamma function \cite{van2002elliptic,spiridonov2008essays,spiridonov2005classical} and $\kappa = (p;p)_\infty (q;q)_\infty $.  Here $a$ is the fugacity for the $SU(2)$ flavor symmetry.

Among other poles in $z$, let us consider the following two at
  \begin{align}\label{polem1m2}
  z= a^{-1} (pq)^{\frac13}  t^{\frac16} q^{m_1}\,,~~~~~~~~~
   z=  (pq)^{\frac13}  t^{-\frac56} q^{-m_2}\,,
  \end{align}
  for some non-negative integers $m_1,m_2$.  These two poles come from   $\Gamma\left(z^{-1}a^{-1} (pq)^{1\over3}t^{1\over6} \right)$ and $\Gamma\left( z(pq)^{-\frac 13 }t^{5\over6}  \right)$,  respectively. 
    This pair of poles collide when
  \begin{align}\label{apole}
  a= t q^{r}\,,~~~~~~r=m_1+m_2\ge0\,.
  \end{align}
At this value of $a$ the contour in $z$ is pinched and produces a simple pole of the $D_3$ index in $a$.   
The residue of  this pole gives the surface defect index with vortex number $r$ in the IR empty $4d$ theory.   The residue can be obtained by first taking the residue of the integrand in \eqref{D3full} with respect to $z$, and then take the residue in $a$ at \eqref{apole}.  The poles in $z$ that contribute to this index  are labeled by a partition $\{m_1,m_2\}$ of the vortex number $r$.


Let us  consider the contribution from the pole in \eqref{polem1m2} labeled by $\{m_1,m_2\}$.  
The residue of the divergent elliptic gamma function is
\begin{align}
&\text{Res}_{z= a^{-1} (pq)^{\frac13}  t^{\frac16} q^{m_1}   }
\, {1\over z}
\Gamma\left( z^{-1} a^{-1} (pq)^{\frac13} t^{1\over6}  \right) 
&={1\over (q;q)(p;p)}\prod_{u=0}^{m_1-1} \prod_{m=0}^\infty  {1\over(  1- p^m q^{-u-1})(1-p^{m+1}q^{u+1})  }\,.
\end{align}
The residue of the integrand at $z= a^{-1} (pq)^{\frac13}   t^{\frac16} q^{m_1}  $ is then
\begin{align}
& \text{Res}_{z= a^{-1} (pq)^{\frac13}  t^{\frac16} q^{m_1}   }
  \, \text{Int}(\mathcal{I}^{D_3})
  ={\kappa\over2} {\Gamma\left(( {pq/t})^{4\over3}  \right) \Gamma\left(  (pq/t)^{1\over3} \right) \over\Gamma\left(( pq/t)^{2\over3} \right)}
{\Gamma\left(a^{-2} pq q^{2m_1} \right) \over 
\Gamma(a^{-2}(pq)^{\frac23} t^{\frac 13}q^{2m_1})}
{\Gamma\left(a^2 (pq)^{-{1\over3}} t^{-\frac23} q^{-2m_1}\right) \over \Gamma(a^{2}(pq)^{-\frac23} t^{-\frac 13}q^{-2m_1})}
\notag\\
&\times
\Gamma\left( (pq)^{\frac23} t^{1\over3} q^{m_1}  \right)
 \Gamma\left( a^2 q^{-m_1} \right) \,
 \Gamma\left(a^{-2} (pq)^{\frac 23} t^{1\over3} q^{m_1}\right)
 \times
 \Gamma\left(a^{-1} (pq)^{2\over3} t^{1\over3} q^{m_1}\right)
 \Gamma\left( a q^{-m_1}\right)\\
 &\times
 \Gamma\left( a^{-1} t q^{m_1}\right)
 \Gamma\left( a(pq)^{-\frac23}  t^{2\over3} q^{-m_1}\right)
   \,  \left( {1\over (q;q)_\infty(p;p)_\infty}\prod_{u=0}^{m_1-1} \prod_{m=0}^\infty  {1\over(  1- p^m q^{-u-1})(1-p^{m+1}q^{u+1})  }\right)\,.\notag
\end{align}
Next, we take the residue of the above expression at $  a= tq^r$ with $r=m_1+m_2$
\begin{align}
\mathcal{A}_{m_1,m_2}&=\text{Res}_{  a= tq^r }\,  {1\over a}
\mathcal{I}^{D_3} 
  =\text{Res}_{  a=  tq^r  }  \,  {1\over a}
\text{Res}_{z= a^{-1} (pq)^{\frac13}  t^{\frac16} q^{m_1}   }
  \, \text{Int}(\mathcal{I}^{D_3})~\notag\\
&  = {1\over2\mathcal{I}_V}
\prod_{m=0}^\infty \left(
\prod_{u=0}^{m_1-1}  {1\over(  1- p^m q^{-u-1})(1-p^{m+1}q^{u+1})  }
\prod_{u=0}^{m_2-1}{1\over(  1- p^m q^{-u-1})(1-p^{m+1}q^{u+1})  }   \right)\notag\\
&\times
 { \Gamma\left( (pq)^{-\frac13}  t^{4\over3} q^{2m_2} \right) \over
\Gamma\left(( {pq})^{-{1\over3}} t^{4\over3}  \right) }
 { \Gamma\left(  (pq/t)^{1\over3} \right)
 \over
 \Gamma\left( (pq/t)^{1\over3}  q^{-m_1}\right)
}
 { \Gamma(pq/t) \over \Gamma\left( (pq/t) q^{-m_2}\right)}
 {\Gamma\left(  (pq/t)^{\frac 23} q^{-m_2}\right)
\over  \Gamma\left(( pq/t)^{2\over3} \right)}\notag\\
&\times
{\Gamma\left(  t^2 q^{m_1+2m_2} \right)
\over\Gamma\left(  t^{2} q^{2m_2}\right) 
}
{
\Gamma\left( (pq)^{-\frac23} t^{\frac53} q^{m_2}\right)
\Gamma\left(  (pq)^{2\over3}  t^{-\frac53} q^{-m_1-2m_2}\right)
  \over
    \Gamma\left(  (pq)^{-\frac23} t^{\frac 53} q^{2m_2} \right)
  \Gamma\left((pq)^{2\over3} t^{-\frac 53} q^{-2m_2} \right)
  }\,,
\end{align}
where
\begin{align}
\mathcal{I}_V = \kappa \Gamma(pq/t)\,
\end{align}
is the superconformal index of an $U(1)$ vector multiplet.   We have used
\begin{align}
\Gamma(z) \Gamma(z^{-1}pq)=1
\end{align}
to simplify the index.

There is one more step we need to do to obtain the surface defect index.   The Higgsed theory in the IR has decoupled  2$d$ sectors that need to be factored out. For the surface defect of vortex number $r$, the decoupled 2$d$ degrees of freedom contribute to the index by
\begin{align}
R_{0,r} = \prod_{u=0}^{r-1} \prod_{m=0}^\infty  {(1-t^{-1} q^{-u}  p^{m+1}  ) (1-tq^u p^m)\over
(1-q^{-u-1}p^m )(1-q^{u+1} p^{m+1})
}\,.
\end{align}
The surface defect index $\mathcal{I}_{\mathbb{S}_r}$ with vortex number $r$ is the residue of the $D_3$ index at $a= t q^{r}$ by summing over the contributions $\mathcal{A}_{m_1,m_2}$ with  the decoupled sectors factored out,
\begin{align}
\mathcal{I}_{\mathbb{S}_r}(p,q,t) = 2\mathcal{I}_V \, R_{0,r}^{-1} \, \text{Res}_{a= t q^{r}  }\,  {1\over a}
\mathcal{I}^{D_3} 
= 2\mathcal{I}_V\,  R_{0,r}^{-1} \,  \sum_{m_1,m_2\ge 0,\, m_1+m_2=r} \mathcal{A}_{m_1,m_2}\,.
\end{align}

\subsection{The Trivial Defect}

Let us consider the case without the defect in an empty $4d$ bulk theory. This $2d$-$4d$ system is completely empty and we expect the index to be 1.  Indeed, the only contribution to the index  comes from the pole labeled by $\{m_1=0,m_2=0\}$, which is
\begin{align}
\mathcal{I}(p,q,t)=1\,,
\end{align}
as expected.

\subsection{The Canonical Surface Defect $r=1$}

Let us move on to consider the $r=1$ vortex defect, i.e. the canonical surface defect, in an empty $4d$ theory.  It is known that the $2d$ theory living on the defect is a twisted Landau-Ginzburg  theory with a cubic twisted superpotential \cite{Gaiotto:2011tf}.  Since the $4d$ theory is empty,  the surface defect index in this case is nothing but the NS-NS elliptic genus of this Landau-Ginzburg theory.

The surface defect index receives contribution from $\{m_1=1,m_2=0\}$ and $\{m_1=0,m_2=1\}$:
\begin{align}
2\mathcal{I}_VR_{0,1}^{-1} \mathcal{A}_{1,0}=&
\prod_{m=0}^\infty
 {1\over(  1- t^{-1} p^m )(1-t p^{m})  }
  { \Gamma\left(  (pq/t)^{1\over3} \right) \over 
\Gamma\left((pq/t)^{1\over3}q^{-1}\right) }
{\Gamma\left(  (pq)^{2\over3}  t^{-\frac53} q^{-1}\right)
  \over
  \Gamma\left((pq)^{2\over3} t^{-\frac 53} \right)
  }
{ \Gamma\left(  t^2 q \right) \over\Gamma\left( t^{2} \right)}\,,
\\
2\mathcal{I}_VR_{0,1}^{-1} \mathcal{A}_{0,1}=&
\prod_{m=0}^\infty
 {1\over(  1- t^{-1} p^m )(1-t p^{m})  }{\Gamma\left( (pq)^{-\frac13} t^{4\over3} q^{2} \right) 
\over
 \Gamma\left(( {pq})^{-{1\over3}}  t^{4\over3} \right)
   }
  {\Gamma\left( t q\right)\over \Gamma(t) }
{\Gamma\left(  (pq/t)^{\frac 23} q^{-1}\right) \over\Gamma\left(( pq/t)^{2\over3} \right)}
{\Gamma\left( (pq)^{-\frac23} t^{\frac53} q\right)
  \over
  \Gamma\left(  (pq)^{-\frac23} t^{\frac 53} q^{2} \right)
  }\notag\,.
\end{align}

Let us simplify the above expression using
\begin{align}
{\Gamma(zq) \over \Gamma(z) }
= (z;p)_\infty (z^{-1} p;p)_\infty 
=\theta_p(z)\,,
\end{align}
where\footnote{Note that $\theta_p(x)$ is related to the Jacobi theta function
\begin{align}
\theta_1 ( p,x)  = - i p^{1\over8} x^{1\over2} \prod_{k=1}^\infty (1-p^k ) (1-xp^k) (1-x^{-1}p^{k-1})\,,
\end{align}
by
\begin{align}
\theta_p (x) = {-i  p^{-\frac18 }x^{\frac12} \over (p;p)_\infty}  \, \theta_1(p,x)\,.
\end{align}}
\begin{align}
\theta_p(x) = {1\over (p;p)_\infty}  \sum_{n\in\mathbb{Z}}p^{n(n-1)/2} (-x)^n\,.
\end{align}

In terms of $\theta_p$, the index can be simplified as 
\begin{align}
&\mathcal{I}_{\mathbb{S}_1}(p,q,t)=
{\theta_p( p^{\frac13 } q^{-\frac23} t^{-\frac13} )   \,
\theta_p(t^2)
\over   \theta_p(p^{\frac23} q^{-\frac13} t^{-\frac53} )  \theta_p(t)}
+
{\theta_p(p^{-\frac13 } q^{-\frac13} t^{\frac43}) \, 
\theta_p (p^{-\frac13} q^{\frac23 } t^{\frac43})\,
\over
\theta_p( p^{\frac23} q^{-\frac13} t^{-\frac23} )\,
\theta_p ( p^{-\frac23 } q^{\frac13} t^{\frac53})
}\,.
\end{align}
Now we want to compare this answer to the elliptic genus of the twisted Landau-Ginzburg model.  To do that, we need to translate the 4$d$ fugacities $p,q,t$ into the 2$d$ fugacities $\mathbf{q},y,e$ in the NS-NS elliptic genus below:
\begin{equation}
\mathcal{G}(\mathbf{q},y,e)=\mathrm{Tr}_{NSNS}\left[(-1)^{F_{2d}}\mathbf{q}^{L_{0}} y^{J_0}\mathbf{\bar{q}}^{\bar{L}_{0}-\bar{J}_{0}/2}e^{C}\right]~,
\end{equation}
where recall that  $e$ is a universal flavor fugacity coupled to the charge $C=R-M_\perp$ \eqref{ccharge}. In this case the twisted Landau-Ginzburg model has no flavor symmetry so the answer should be independent of $e$.  
The $4d$ fugacities are related to the $2d$ ones as (see, for example, \cite{Gadde:2013ftv,Cordova:2017ohl})
\begin{align}
p=\mathbf{q} \,,~~~~q=\mathbf{q}^{1\over2} e y\,,~~~~~
t=\mathbf{q} e y^2\,.
\end{align}
  
In terms of the 2$d$ fugacities, the surface defect index becomes
\begin{align}
&\mathcal{I}_{\mathbb{S}_1}(p,q,t)=
{\theta_\mathbf{q}( \mathbf{q}^{-\frac13 } y^{-\frac 43} e^{-1} )   \,
\theta_\mathbf{q} (\mathbf{q}^2 y^4 e^2)
\over   \theta_\mathbf{q} (\mathbf{q}^{-\frac76}  y^{-\frac{11}{3}} e^{-2})
\theta_{\mathbf{q}} (\mathbf{q} y^2e)
}
+
{\theta_\mathbf{q} (\mathbf{q}^{\frac56 }y^{\frac 73} e) \, 
\theta_\mathbf{q} ( \mathbf{q}^{\frac43} y^{10\over3}  e^2 )\,
\over
\theta_\mathbf{q} (\mathbf{q}^{-\frac16} y^{-\frac53} e^{-1})\,
\theta_\mathbf{q}  ( \mathbf{q}^{\frac 76} y^{11\over3}  e^2)
}\,.
\end{align}
One can check that the above answer has the same series expansion as the  elliptic genus of the twisted Landau-Ginzburg model with cubic superpotential \cite{Witten:1993jg},
\begin{align}
\mathcal{G}(\mathbf{q},y)
=y\mathbf{q}^{1\over2}  {\theta_\mathbf{q}(\mathbf{q}^{-\frac23}y^{-\frac23 })\over \theta_\mathbf{q}( \mathbf{q}^{-\frac16} y^{\frac13})}\,.
\end{align}
In particular, the answer in this case does not depend on the universal flavor fugacity $e$.  We have thus obtained the expected full superconformal index for the canonical surface defect in the empty $4d$ theory from Higgsing $D_3$ Argyres-Douglas theory.

\subsection{The Vortex Defect with $r=2$}

For the vortex number $r=2$ defect, we saw that the Schur index is zero in \eqref{emptyschur}, which is consistent with the fact that the chiral algebra is trivial.  Here we will see that it  has a nontrivial Macdonald index.  

In this case we need to add up the contributions from $\{m_1=2,m_2=0\}$, $\{m_1=1,m_2=1\}$, and $\{m_1=0,m_2=2\}$. For simplicity, we will only compute the Macdonald limit of the index.  The three contributions in this limit become
\begin{align}
\begin{split}
&2\mathcal{I}_VR_{0,2}^{-1} \mathcal{A}_{2,0}= {(1+t)(1-t^2q)\over (1-tq)}\,,\\
&2\mathcal{I}_VR_{0,2}^{-1} \mathcal{A}_{1,1}=- t  (1+ tq){1-q^{-2}\over 1-q^{-1}}\,,\\
&2\mathcal{I}_VR_{0,2}^{-1} \mathcal{A}_{0,2}= t^2 q\,.
\end{split}
\end{align}
Adding all these contributions together, we obtain the $r=2$ vortex defect index in an empty $4d$ theory in the Macdonald limit,
  \begin{align}
  \mathcal{I}_{\mathbb{S}_2}(q,t) = {1-tq^{-1} \over 1-q t} \,.
  \end{align}
  Indeed it vanishes in the Schur limit $t=q$.  The Macdonald index takes the form of a character with one bosonic mode and one fermionic mode.  It is intriguing that even though the chiral algebra is trivial in this case, there is still a nontrivial Macdonald index.

\subsection{The Vortex Defect with $r=3$}


For the $r=3$ vortex defect index, there are four contributions from the poles labeled by $\{m_1=3,m_2=0\}\,, \{m_1=2,m_2=1\},, \{m_1=1,m_2=2\}\,, \{m_1=0,m_2=3\}$.  Adding all the above together, we obtain the Macdonald index for the $r=3$ vortex defect,
\begin{align}
\mathcal{I}_{\mathbb{S}_3}(q,t) = \frac{q^3 t+q^2-q t-t}{q^2-q^4 t}\,.
\end{align}
This reduces to $-1/q$ in the $t=q$ Schur limit. Note that the Macdonald index does not enjoy the periodicity $\bar r \leftrightarrow 1-\bar r$ of the Schur index, where $\bar r=r~\text{mod}~3$.

\section{The Full Superconformal Index for Vortex Defects of the Hypermultiplet}
\label{sec:fullhyper}

In this appendix we extend the computation of the full, three-variable surface defect index in the previous appendix to  the free $4d$ hypermultiplet theory. Even though the $4d$ bulk theory is free, the coupled $2d$-$4d$ system is generally strongly interacting and has   interesting nontrivial defect indices as we saw in section \ref{sec:freehyper}.

 We again  focus on vortex defects $\mathbb{S}_r$ arising from turning on a position-dependent Higgs field in the $D_4$ Argyres-Douglas theory, and then flowing to the IR free hypermultiplet theory with surface defects.  As discussed in section \ref{sec:higgs},  the surface defect index of a free hypermultiplet can be computed from the residues of the index for the $D_4$ theory without the defect.

Let us motivate our consideration of superconformal indices beyond the Schur limit. First, recall that in \eqref{IRhyper}, we saw that the canonical surface defect (i.e. vortex number $r=1$)  is naively 0.  We would like to understand whether this implies that the spectrum of the 2$d$-4$d$ supersymmetric operators is empty, or there is an infinite degeneracy leading to 0 after certain regularization.  As discussed before, the correspondence with the chiral algebra  suggests the latter interpretation is the correct one.  In this appendix we  explicit compute the full superconformal index of the canonical surface defect and show that it is nonzero, and thus confirming the picture from the chiral algebra.

Secondly, for the surface defect with vortex number $r=2$, its Schur index was shown to agree with the original Schur index of the free hypermultiplet without  defects.  We would like to verify that this $r=2$ vortex defect is physically distinct from the trivial defect, even though they share the same spectrum of 2$d$-4$d$ Schur operators.  In the following we will compute the full superconformal index of this $r=2$ vortex defect and show that it differs from the index of the free hypermultiplet in the absence of defects.  These analyses give a concrete example of two physically distinct surface defects sharing the same 2$d$-4$d$ Schur operator spectrum, and hence correspond to the same module in the associated chiral algebra.

Let us embark on the calculation.  The full superconformal index of the $D_4$ theory has been conjectured to be \cite{Maruyoshi:2016aim,Agarwal:2016pjo},
\begin{align}\label{D4full}
&\mathcal{I}^{D_4}  (p,q,t,a_1,a_2) 
={\kappa\over2} {\Gamma\left(( {pq/t})^{3\over2}  \right) \Gamma\left(  (pq/t)^{1\over2} \right) \over\Gamma\left( pq/t \right)}
\oint {dz\over 2\pi i z} {\Gamma\left(z^{\pm 2} (pq)^{1\over2} t^{-\frac12} \right) \over \Gamma(z^{\pm2})}\notag\\
&\times
 \Gamma\left((za_1a_2^3)^\pm (pqt)^{1\over4} \right) \,
 \Gamma\left((z^{-1} a_1a_2^3)^\pm (pqt)^{1\over4}\right)
\,  \Gamma\left((za_1a_2^{-1})^\pm (pq)^{-\frac14} t^{3\over4} \right) \, 
\Gamma\left(  (z^{-1} a_1a_2^{-1})^\pm (pq)^{-{1\over4}} t^{3\over4}\right)\,,
\end{align}
where $a_1$ and $a_2$ are the fugacities for the $SU(3)$ flavor symmetry.  Incidentally, $\Gamma(z)$ is  the superconformal index of a half-hypermultiplet.

Among other poles in $z$, let us consider the following two
\begin{align}\label{m1m2pole}
  z= a_1a_2^{3} (pq)^{-\frac14}  t^{-\frac14} q^{-m_1}\,,~~~~~~~~~
  z=a_1 a_2^{-1}  (pq)^{-\frac14} t^{3\over4}q^{m_2}\,,
  \end{align}
  for some non-negative integers $m_1,m_2$. They come from  $\Gamma\left((z^{-1} a_1a_2^3)^{-1} (pqt)^{1\over4}\right)$ and \\$\Gamma\left(  (z^{-1} a_1a_2^{-1}) (pq)^{-{1\over4}} t^{3\over4}\right)$, respectively.  
 This pair of poles collide when
  \begin{align}
  a_2 = t^{1\over4 } q^{r\over 4}\,,~~~~~~r=m_1+m_2\ge0\,.
  \end{align}
  At this value of $a_2$ the contour in $z$ is pinched and produce a simple pole of the $D_4$ index in $a_2$.  
As in the previous appendix, this residue for the $D_4$ index  gives the index of the surface defect of vortex number $r$ in the free hypermultiplet theory.   


Let us  consider the contribution from the pole \eqref{m1m2pole} in $z$ labeled by $\{m_1,m_2\}$.  A similar calculation as in the previous appendix shows that the contribution to the residue of the $D_4$ index at $a_2 = t^{1\over4 } q^{r\over4}$ from the pole $\{m_1,m_2\}$ is
\begin{align}
& \mathcal{A}_{m_1,m_2} =\text{Res}_{a_2^2= t^{1\over2}  q^{r\over2}  }  \,  {1\over a_2}
\text{Res}_{z= a_1a_2^{-1} (pq)^{-\frac14} t^{3\over4} q^{m_2}   }
  \, \text{Int}(\mathcal{I}^{D_4})~\notag\\
&  = {1\over2\mathcal{I}_V}
\prod_{m=0}^\infty \left(
\prod_{u=0}^{m_1-1}  {1\over(  1- p^m q^{-u-1})(1-p^{m+1}q^{u+1})  }
\prod_{u=0}^{m_2-1}{1\over(  1- p^m q^{-u-1})(1-p^{m+1}q^{u+1})  }   \right)\notag\\
& {  \Gamma\left(( {pq/t})^{3\over2}  \right) \,
   \Gamma\left(  (pq/t)^{1\over2} \right)  }
\times 
{ \Gamma\left(  a_1^2 t^{1\over2}  {  q^{-m_1 +3m_2\over 2} }\right) \, 
  \Gamma\left(  a_1^{-2}  pq t^{-\frac32}  {  q^{m_1-3m_2\over2} }  \right)
  \over
\Gamma\left( a_1^2 (pq)^{-\frac12}  t    {  q^{-m_1+3m_2 \over2 }}    \right)  \, 
 \Gamma \left( a_1^{-2}  (pq)^{1\over2}  t^{-1}   {  q^{m_1-3m_2\over2}}   \right) }\notag\\
& \times    \Gamma\left( a_1^2 t^{3\over2}   {  q^{m_1+3m_2\over2  } } \right)   \,
 \Gamma\left( a_1^{-2}  (pq)^{1\over2}  t^{-1}  {  q^{-m_1-3m_2\over2 } } \right)  \,
  \Gamma\left(  (pq)^{1\over2} t^{1\over2}    {  q^{m_1 } }   \right)     \notag\\
 &\times    \Gamma\left( a_1^2  (pq)^{-\frac12}  t    {  q^{-m_1+m_2\over2 } }   \right)   \,
 \Gamma\left(a_1^{-2} t^{1\over2}   {  q^{m_1-m_2\over2 } }\right) \,
  \Gamma\left(   (pq)^{-\frac12}  t^{3\over2}  {  q^{m_2  } } \right)  \,.
\end{align}

\subsection{The Trivial Defect}

Let us test the above formula by considering the case without surface defects, i.e. $r=m_1=m_2=0$. We find
\begin{align}
\mathcal{I}(p,q,t,a_1) = 2\mathcal{I}_V \mathcal{A}_{0,0} = \Gamma(a_1^2 t^{1\over2}) \Gamma(a_1^{-2} t^{1\over2})\,,
\end{align}
which is indeed the superconformal index of a free hypermultiplet.  

\subsection{The Canonical Surface Defect $r=1$}

Let us consider the case of the canonical surface defect, i.e. $r=1$. The full defect index can be simplified to\footnote{If we directly take the Schur limit $t=q$, the surface defect index is not analytic in $q$ and seems to depend on $p$.  This is most likely due to the infinite degeneracy of Schur operators mentioned previously. We believe the direct  Schur limit can be analytically continued to 0.}
\begin{align}\label{fullhyperindapend}
&\mathcal{I}_{\mathbb{S}_1}(p,q,t,a_1)
= \Gamma(a_1^{\pm2} q^{-\frac12} t^{\frac12}  )
{1\over    \theta_p(t) \theta_p(a_1^2 t) }\\
&\times\left[  \,
\theta_p(p^{1\over2} q^{-\frac12} t^{1\over2} )   
\theta_p(a_1^2 q^{-\frac12 } t^{\frac32} )
\theta_p (a_1^2 pq^{\frac12} t^{-\frac12})
+ a_1^{-2} q^{-\frac12} t^{1\over2} \,
\theta_p(p^{1\over2} q^{-\frac12} t^{\frac32} )
\theta_p(a_1^2 q^{-\frac12} t^{\frac12})
\theta_p(a_1^2 q^{\frac12} t^{\frac12})
\,
\right]\notag\,.
\end{align}
Let us consider the Macdonald limit \eqref{macdonald} $p\to0$ of the canonical surface defect in a free hypermultiplet theory:
\begin{align}\label{Macdonald1}
&\mathcal{I}_{\mathbb{S}_1}(q,t,a_1)
  ={1-tq^{-1}\over  (a_1^2 t^{1\over2} q^{-\frac12} ;q)_\infty (a_1^{-2}t^{1\over2} q^{-{1\over2}};q)_\infty } \,.
\end{align}
The Schur index is a further limit of the Macdonald index by setting $t=q$, which gives 0 we already saw in \eqref{IRhyper}.  

For a general 4$d$ SCFT  with discrete spectrum, the Macdonald index admits  an expansion in $q$  while keeping $t/q$ fixed.  If we perform such an expansion for \eqref{Macdonald1}, we find that at each power of $q$ there are infinitely many terms coming from the two $q$-Pochhammer symbols.  This is consistent with the infinite degeneracy at each level in the twisted module of the corresponding chiral algebra.

\subsection{The Vortex Defect with $r=2$}

More generally, the contribution from the $\{m_1,m_2\}$ pole to the  Macdonald index is
\begin{align}
&2\mathcal{I}_V\mathcal{A}_{m_1,m_2}\\
&= \left(\prod_{u=0}^{m_1-1} {1\over 1-q^{-u-1}}
\prod_{u=0}^{m_2-1} {1\over 1-q^{-u-1}}\right)
{a_1^{-2m_2} t^{m_2\over2} q^{(m_1-m_2)m_2\over 2}(a_1^2 t^{3\over2} q^{- m_1+3m_2 \over2}  ;q)_\infty \over
(a_1^2 t^{1\over2} q^{-m_1+3m_2\over2} ;q)_\infty
(a_1^2 t^{3\over2} q^{m_1+3m_2\over2} ;q)_\infty  
(a_1^{-2} t^{1\over2} q^{m_1-m_2\over2};q)_\infty}\,.\notag
\end{align}
The Macdonald index of the $r=2$ vortex defect $\mathbb{S}_2$  is then
\begin{align}
&\mathcal{I}_{\mathbb{S}_2} (q,t,a_1) = 2\mathcal{I}_V\, R_{0,2}^{-1}\, (\mathcal{A}_{2,0} + \mathcal{A}_{1,1}+\mathcal{A}_{0,2}) \\
&=   
\Big[\, 
{ (a_1^2+ a_1^{-2})t^{3\over2} (1-q^2)  - (t+2qt -2qt^2-q^2 +t^3  -q^2t^2)
\over
(1-tq)(a_1^2 q-t^{1\over2}) (a_1^{-2} q - t^{1\over2} )}
  \Big]
  {1\over (a_1^2t^{1\over2} ;q)_\infty (a_1^{-2} t^{1\over2} ;q)_\infty}
 \notag\,.
\end{align}
The rightmost factor ${1\over (a_1^2t^{1\over2} ;q)_\infty (a_1^{-2} t^{1\over2} ;q)_\infty}
$ is the Macdonald index for a free hypermultiplet.  In the Schur limit $t=q$, we recover, up to an overall sign, the Schur index of the free hypermultiplet without defects,
\begin{align}
- \,  {1\over (a_1^2  q^{1\over2} ;q)_\infty (a_1^{-2} q^{1\over2} ;q)_\infty} \,.
\end{align}
We conclude that even though the $r=2$ vortex defect $\mathbb{S}_2$ shares the same Schur index and the trivial defect in the free hypermultiplets, they are in fact physically distinct and can be already distinguished by the more refined Macdonald index.

\bibliography{2d4dAD}{}

\providecommand{\href}[2]{#2}\begingroup\raggedright\begin{thebibliography}{10}

\bibitem{Kinney:2005ej}
J.~Kinney, J.~M. Maldacena, S.~Minwalla, and S.~Raju, ``{An Index for 4
  dimensional super conformal theories},'' {\em Commun. Math. Phys.} {\bf 275}
  (2007) 209--254,
\href{http://www.arXiv.org/abs/hep-th/0510251}{{\tt hep-th/0510251}}.

\bibitem{Gadde:2011ik}
A.~Gadde, L.~Rastelli, S.~S. Razamat, and W.~Yan, ``{The 4d Superconformal
  Index from q-deformed 2d Yang-Mills},'' {\em Phys. Rev. Lett.} {\bf 106}
  (2011) 241602,
\href{http://www.arXiv.org/abs/1104.3850}{{\tt 1104.3850}}.

\bibitem{Gadde:2011uv}
A.~Gadde, L.~Rastelli, S.~S. Razamat, and W.~Yan, ``{Gauge Theories and
  Macdonald Polynomials},'' {\em Commun.Math.Phys.} {\bf 319} (2013) 147--193,
\href{http://www.arXiv.org/abs/1110.3740}{{\tt 1110.3740}}.

\bibitem{Gadde:2009kb}
A.~Gadde, E.~Pomoni, L.~Rastelli, and S.~S. Razamat, ``{S-duality and 2d
  Topological QFT},'' {\em JHEP} {\bf 1003} (2010) 032,
\href{http://www.arXiv.org/abs/0910.2225}{{\tt 0910.2225}}.

\bibitem{Beem:2013sza}
C.~Beem, M.~Lemos, P.~Liendo, W.~Peelaers, L.~Rastelli, {\em et al.},
  ``{Infinite Chiral Symmetry in Four Dimensions},'' {\em Commun.Math.Phys.}
  {\bf 336} (2015), no.~3, 1359--1433,
\href{http://www.arXiv.org/abs/1312.5344}{{\tt 1312.5344}}.

\bibitem{Iqbal:2012xm}
A.~Iqbal and C.~Vafa, ``{BPS Degeneracies and Superconformal Index in Diverse
  Dimensions},'' {\em Phys. Rev.} {\bf D90} (2014), no.~10, 105031,
\href{http://www.arXiv.org/abs/1210.3605}{{\tt 1210.3605}}.

\bibitem{Cordova:2015nma}
C.~C\'ordova and S.-H. Shao, ``{Schur Indices, BPS Particles, and
  Argyres-Douglas Theories},'' {\em JHEP} {\bf 01} (2016) 040,
\href{http://www.arXiv.org/abs/1506.00265}{{\tt 1506.00265}}.

\bibitem{Gukov:2014gja}
S.~Gukov, ``{Surface Operators},'' in {\em New Dualities of Supersymmetric
  Gauge Theories}, J.~Teschner, ed., pp.~223--259.
\newblock 2016.
\newblock
\href{http://www.arXiv.org/abs/1412.7127}{{\tt 1412.7127}}.
\newblock

\bibitem{Beem:2014rza}
C.~Beem, W.~Peelaers, L.~Rastelli, and B.~C. van Rees, ``{Chiral algebras of
  class S},'' {\em JHEP} {\bf 1505} (2015) 020,
\href{http://www.arXiv.org/abs/1408.6522}{{\tt 1408.6522}}.

\bibitem{Lemos:2014lua}
M.~Lemos and W.~Peelaers, ``{Chiral Algebras for Trinion Theories},'' {\em
  JHEP} {\bf 1502} (2015) 113,
\href{http://www.arXiv.org/abs/1411.3252}{{\tt 1411.3252}}.

\bibitem{Liendo:2015ofa}
P.~Liendo, I.~Ramirez, and J.~Seo, ``{Stress-tensor OPE in $ \mathcal{N}=2 $
  superconformal theories},'' {\em JHEP} {\bf 02} (2016) 019,
\href{http://www.arXiv.org/abs/1509.00033}{{\tt 1509.00033}}.

\bibitem{Lemos:2015orc}
M.~Lemos and P.~Liendo, ``{$\mathcal{N}=2$ central charge bounds from $2d$
  chiral algebras},'' {\em JHEP} {\bf 04} (2016) 004,
\href{http://www.arXiv.org/abs/1511.07449}{{\tt 1511.07449}}.

\bibitem{arakawa2015joseph}
T.~Arakawa and A.~Moreau, ``{Joseph ideals and lisse minimal W-algebras},''
\href{http://www.arXiv.org/abs/1506.00710}{{\tt 1506.00710}}.

\bibitem{Nishinaka:2016hbw}
T.~Nishinaka and Y.~Tachikawa, ``{On 4d rank-one N=3 superconformal field
  theories},''
\href{http://www.arXiv.org/abs/1602.01503}{{\tt 1602.01503}}.

\bibitem{Buican:2016arp}
M.~Buican and T.~Nishinaka, ``{Conformal Manifolds in Four Dimensions and
  Chiral Algebras},''
\href{http://www.arXiv.org/abs/1603.00887}{{\tt 1603.00887}}.

\bibitem{Arakawa:2016hkg}
T.~Arakawa and K.~Kawasetsu, ``{Quasi-lisse vertex algebras and modular linear
  differential equations},''
\href{http://www.arXiv.org/abs/1610.05865}{{\tt 1610.05865}}.

\bibitem{Bonetti:2016nma}
F.~Bonetti and L.~Rastelli, ``{Supersymmetric Localization in AdS$_5$ and the
  Protected Chiral Algebra},''
\href{http://www.arXiv.org/abs/1612.06514}{{\tt 1612.06514}}.

\bibitem{beem}
C.~Beem and L.~Rastelli, ``{Vertex Operators, Higgs Branches, and Modular
  Differential Equations},'' {\em to appear}.

\bibitem{BPR}
C.~Beem, W.~Peelaers, and L.~Rastelli, work in~progress.

\bibitem{Gaiotto:2012xa}
D.~Gaiotto, L.~Rastelli, and S.~S. Razamat, ``{Bootstrapping the superconformal
  index with surface defects},'' {\em JHEP} {\bf 01} (2013) 022,
\href{http://www.arXiv.org/abs/1207.3577}{{\tt 1207.3577}}.

\bibitem{Hanany:1997vm}
A.~Hanany and K.~Hori, ``{Branes and N=2 theories in two-dimensions},'' {\em
  Nucl. Phys.} {\bf B513} (1998) 119--174,
\href{http://www.arXiv.org/abs/hep-th/9707192}{{\tt hep-th/9707192}}.

\bibitem{Gaiotto:2009fs}
D.~Gaiotto, ``{Surface Operators in N = 2 4d Gauge Theories},'' {\em JHEP} {\bf
  11} (2012) 090,
\href{http://www.arXiv.org/abs/0911.1316}{{\tt 0911.1316}}.

\bibitem{Gaiotto:2011tf}
D.~Gaiotto, G.~W. Moore, and A.~Neitzke, ``{Wall-Crossing in Coupled 2d-4d
  Systems},'' {\em JHEP} {\bf 12} (2012) 082,
\href{http://www.arXiv.org/abs/1103.2598}{{\tt 1103.2598}}.

\bibitem{Cordova:2017ohl}
C.~C\'{o}rdova, D.~Gaiotto, and S.-H. Shao, ``{Surface Defect Indices and 2d-4d
  BPS States},''
\href{http://www.arXiv.org/abs/1703.02525}{{\tt 1703.02525}}.

\bibitem{Argyres:1995jj}
P.~C. Argyres and M.~R. Douglas, ``{New phenomena in SU(3) supersymmetric gauge
  theory},'' {\em Nucl.Phys.} {\bf B448} (1995) 93--126,
\href{http://www.arXiv.org/abs/hep-th/9505062}{{\tt hep-th/9505062}}.

\bibitem{Argyres:1995xn}
P.~C. Argyres, M.~R. Plesser, N.~Seiberg, and E.~Witten, ``{New N=2
  superconformal field theories in four-dimensions},'' {\em Nucl.Phys.} {\bf
  B461} (1996) 71--84,
\href{http://www.arXiv.org/abs/hep-th/9511154}{{\tt hep-th/9511154}}.

\bibitem{Shapere:1999xr}
A.~D. Shapere and C.~Vafa, ``{BPS structure of Argyres-Douglas superconformal
  theories},''
\href{http://www.arXiv.org/abs/hep-th/9910182}{{\tt hep-th/9910182}}.

\bibitem{Gaiotto:2009hg}
D.~Gaiotto, G.~W. Moore, and A.~Neitzke, ``{Wall-crossing, Hitchin Systems, and
  the WKB Approximation},''
\href{http://www.arXiv.org/abs/0907.3987}{{\tt 0907.3987}}.

\bibitem{Gaiotto:2010be}
D.~Gaiotto, G.~W. Moore, and A.~Neitzke, ``{Framed BPS States},'' {\em Adv.
  Theor. Math. Phys.} {\bf 17} (2013), no.~2, 241--397,
\href{http://www.arXiv.org/abs/1006.0146}{{\tt 1006.0146}}.

\bibitem{Cecotti:2010fi}
S.~Cecotti, A.~Neitzke, and C.~Vafa, ``{R-Twisting and 4d/2d
  Correspondences},''
\href{http://www.arXiv.org/abs/1006.3435}{{\tt 1006.3435}}.

\bibitem{Alim:2011ae}
M.~Alim, S.~Cecotti, C.~C\'ordova, S.~Espahbodi, A.~Rastogi, and C.~Vafa,
  ``{BPS Quivers and Spectra of Complete N=2 Quantum Field Theories},'' {\em
  Commun. Math. Phys.} {\bf 323} (2013) 1185--1227,
\href{http://www.arXiv.org/abs/1109.4941}{{\tt 1109.4941}}.

\bibitem{Alim:2011kw}
M.~Alim, S.~Cecotti, C.~C\'ordova, S.~Espahbodi, A.~Rastogi, {\em et al.},
  ``{$\mathcal{N} = 2$ quantum field theories and their BPS quivers},'' {\em
  Adv.Theor.Math.Phys.} {\bf 18} (2014) 27--127,
\href{http://www.arXiv.org/abs/1112.3984}{{\tt 1112.3984}}.

\bibitem{Maruyoshi:2013fwa}
K.~Maruyoshi, C.~Y. Park, and W.~Yan, ``{BPS spectrum of Argyres-Douglas theory
  via spectral network},'' {\em JHEP} {\bf 12} (2013) 092,
\href{http://www.arXiv.org/abs/1309.3050}{{\tt 1309.3050}}.

\bibitem{Cordova:2013bza}
C.~C\'ordova and A.~Neitzke, ``{Line Defects, Tropicalization, and
  Multi-Centered Quiver Quantum Mechanics},'' {\em JHEP} {\bf 09} (2014) 099,
\href{http://www.arXiv.org/abs/1308.6829}{{\tt 1308.6829}}.

\bibitem{Beem:2014zpa}
C.~Beem, M.~Lemos, P.~Liendo, L.~Rastelli, and B.~C. van Rees, ``{The $
  \mathcal{N}=2 $ superconformal bootstrap},'' {\em JHEP} {\bf 03} (2016) 183,
\href{http://www.arXiv.org/abs/1412.7541}{{\tt 1412.7541}}.

\bibitem{Buican:2015ina}
M.~Buican and T.~Nishinaka, ``{On the superconformal index of Argyres-Douglas
  theories},'' {\em J. Phys.} {\bf A49} (2016), no.~1, 015401,
\href{http://www.arXiv.org/abs/1505.05884}{{\tt 1505.05884}}.

\bibitem{Xie:2016evu}
D.~Xie, W.~Yan, and S.-T. Yau, ``{Chiral algebra of Argyres-Douglas theory from
  M5 brane},''
\href{http://www.arXiv.org/abs/1604.02155}{{\tt 1604.02155}}.

\bibitem{Creutzig:2017qyf}
T.~Creutzig, ``{W-algebras for Argyres-Douglas theories},''
\href{http://www.arXiv.org/abs/1701.05926}{{\tt 1701.05926}}.

\bibitem{Song:2015wta}
J.~Song, ``{Superconformal indices of generalized Argyres-Douglas theories from
  2d TQFT},'' {\em JHEP} {\bf 02} (2016) 045,
\href{http://www.arXiv.org/abs/1509.06730}{{\tt 1509.06730}}.

\bibitem{Cecotti:2015lab}
S.~Cecotti, J.~Song, C.~Vafa, and W.~Yan, ``{Superconformal Index, BPS
  Monodromy and Chiral Algebras},''
\href{http://www.arXiv.org/abs/1511.01516}{{\tt 1511.01516}}.

\bibitem{Cordova:2016uwk}
C.~C\'ordova, D.~Gaiotto, and S.-H. Shao, ``{Infrared Computations of Defect
  Schur Indices},'' {\em JHEP} {\bf 11} (2016) 106,
\href{http://www.arXiv.org/abs/1606.08429}{{\tt 1606.08429}}.

\bibitem{Fredrickson:2017yka}
L.~Fredrickson, D.~Pei, W.~Yan, and K.~Ye, ``{Argyres-Douglas Theories, Chiral
  Algebras and Wild Hitchin Characters},''
\href{http://www.arXiv.org/abs/1701.08782}{{\tt 1701.08782}}.

\bibitem{Buican:2015tda}
M.~Buican and T.~Nishinaka, ``{Argyres-Douglas Theories, the Macdonald Index,
  and an RG Inequality},'' {\em JHEP} {\bf 02} (2016) 159,
\href{http://www.arXiv.org/abs/1509.05402}{{\tt 1509.05402}}.

\bibitem{Maruyoshi:2016tqk}
K.~Maruyoshi and J.~Song, ``{The Full Superconformal Index of the
  Argyres-Douglas Theory},''
\href{http://www.arXiv.org/abs/1606.05632}{{\tt 1606.05632}}.

\bibitem{Maruyoshi:2016aim}
K.~Maruyoshi and J.~Song, ``{N=1 Deformations and RG Flows of N=2 SCFTs},''
\href{http://www.arXiv.org/abs/1607.04281}{{\tt 1607.04281}}.

\bibitem{Agarwal:2016pjo}
P.~Agarwal, K.~Maruyoshi, and J.~Song, ``{$ \mathcal{N} $ =1 Deformations and
  RG flows of $ \mathcal{N} $ =2 SCFTs, part II: non-principal deformations},''
  {\em JHEP} {\bf 12} (2016) 103,
\href{http://www.arXiv.org/abs/1610.05311}{{\tt 1610.05311}}.

\bibitem{Nakayama:2011pa}
Y.~Nakayama, ``{4D and 2D superconformal index with surface operator},'' {\em
  JHEP} {\bf 08} (2011) 084,
\href{http://www.arXiv.org/abs/1105.4883}{{\tt 1105.4883}}.

\bibitem{Gadde:2013ftv}
A.~Gadde and S.~Gukov, ``{2d Index and Surface operators},'' {\em JHEP} {\bf
  03} (2014) 080,
\href{http://www.arXiv.org/abs/1305.0266}{{\tt 1305.0266}}.

\bibitem{Benini:2013nda}
F.~Benini, R.~Eager, K.~Hori, and Y.~Tachikawa, ``{Elliptic genera of
  two-dimensional N=2 gauge theories with rank-one gauge groups},'' {\em Lett.
  Math. Phys.} {\bf 104} (2014) 465--493,
\href{http://www.arXiv.org/abs/1305.0533}{{\tt 1305.0533}}.

\bibitem{Benini:2013xpa}
F.~Benini, R.~Eager, K.~Hori, and Y.~Tachikawa, ``{Elliptic Genera of 2d
  ${\mathcal{N}}$ = 2 Gauge Theories},'' {\em Commun. Math. Phys.} {\bf 333}
  (2015), no.~3, 1241--1286,
\href{http://www.arXiv.org/abs/1308.4896}{{\tt 1308.4896}}.

\bibitem{Cecotti:1992rm}
S.~Cecotti and C.~Vafa, ``{On classification of N=2 supersymmetric theories},''
  {\em Commun.Math.Phys.} {\bf 158} (1993) 569--644,
\href{http://www.arXiv.org/abs/hep-th/9211097}{{\tt hep-th/9211097}}.

\bibitem{Gaiotto:2015aoa}
D.~Gaiotto, G.~W. Moore, and E.~Witten, ``{Algebra of the Infrared: String
  Field Theoretic Structures in Massive ${\cal N}=(2,2)$ Field Theory In Two
  Dimensions},''
\href{http://www.arXiv.org/abs/1506.04087}{{\tt 1506.04087}}.

\bibitem{Seiberg:1994rs}
N.~Seiberg and E.~Witten, ``{Electric - magnetic duality, monopole
  condensation, and confinement in N=2 supersymmetric Yang-Mills theory},''
  {\em Nucl. Phys.} {\bf B426} (1994) 19--52,
  \href{http://www.arXiv.org/abs/hep-th/9407087}{{\tt hep-th/9407087}}.
[Erratum: Nucl. Phys.B430,485(1994)].

\bibitem{Seiberg:1994aj}
N.~Seiberg and E.~Witten, ``{Monopoles, duality and chiral symmetry breaking in
  N=2 supersymmetric QCD},'' {\em Nucl. Phys.} {\bf B431} (1994) 484--550,
\href{http://www.arXiv.org/abs/hep-th/9408099}{{\tt hep-th/9408099}}.

\bibitem{CD}
C.~C\'{o}rdova and T.~Dumitrescu, ``{Current Algebra Constraints on BPS
  Particles},'' {\em to appear}.

\bibitem{drinfeld1984lie}
V.~G. Drinfeld and V.~V. Sokolov, ``{Lie Algebras and Equations of Korteweg-de
  Vries Type},'' {\em Itogi Nauki i Tekhniki. Seriya" Sovremennye Problemy
  Matematiki. Noveishie Dostizheniya"} {\bf 24} (1984) 81--180.

\bibitem{deBoer:1992sy}
J.~de~Boer and T.~Tjin, ``{Quantization and representation theory of finite W
  algebras},'' {\em Commun. Math. Phys.} {\bf 158} (1993) 485--516,
\href{http://www.arXiv.org/abs/hep-th/9211109}{{\tt hep-th/9211109}}.

\bibitem{deBoer:1993iz}
J.~de~Boer and T.~Tjin, ``{The Relation between quantum W algebras and Lie
  algebras},'' {\em Commun. Math. Phys.} {\bf 160} (1994) 317--332,
\href{http://www.arXiv.org/abs/hep-th/9302006}{{\tt hep-th/9302006}}.

\bibitem{Lesage:2002ch}
F.~Lesage, P.~Mathieu, J.~Rasmussen, and H.~Saleur, ``{The
  $\widehat{su}(2)_{-1/2}$ WZW model and the beta gamma system},'' {\em Nucl.
  Phys.} {\bf B647} (2002) 363--403,
\href{http://www.arXiv.org/abs/hep-th/0207201}{{\tt hep-th/0207201}}.

\bibitem{Song:2016yfd}
J.~Song, ``{Macdonald Index and Chiral Algebra},''
\href{http://www.arXiv.org/abs/1612.08956}{{\tt 1612.08956}}.

\bibitem{philippe1997conformal}
P.~D. Francesco, P.~Mathieu, and D.~Senechal, {\em Conformal Field Theory}.
\newblock Graduate Texts in Contemporary Physics. Springer, 1997.

\bibitem{Feigin:2004wb}
B.~L. Feigin and A.~M. Semikhatov, ``{W(2)(n) algebras},'' {\em Nucl. Phys.}
  {\bf B698} (2004) 409--449,
\href{http://www.arXiv.org/abs/math/0401164}{{\tt math/0401164}}.

\bibitem{Creutzig:2013pda}
T.~Creutzig, D.~Ridout, and S.~Wood, ``{Coset Constructions of Logarithmic (1,
  p) Models},'' {\em Lett. Math. Phys.} {\bf 104} (2014) 553--583,
\href{http://www.arXiv.org/abs/1305.2665}{{\tt 1305.2665}}.

\bibitem{Bershadsky:1990bg}
M.~Bershadsky, ``{Conformal field theories via Hamiltonian reduction},'' {\em
  Commun. Math. Phys.} {\bf 139} (1991)
71--82.

\bibitem{Polyakov:1989dm}
A.~M. Polyakov, ``{Gauge Transformations and Diffeomorphisms},'' {\em Int. J.
  Mod. Phys.} {\bf A5} (1990)
833.

\bibitem{Whalen:2014fta}
D.~Whalen, ``{An algorithm for evaluating Gram matrices in Verma modules of
  W-algebras},''
\href{http://www.arXiv.org/abs/1412.0759}{{\tt 1412.0759}}.

\bibitem{van2002elliptic}
J.~Van~Diejen and V.~Spiridonov, ``{Elliptic Beta Integrals and Modular
  Hypergeometric Sums: An Overview},'' {\em Journal of Mathematics} {\bf 32}
  (2002), no.~2,.

\bibitem{spiridonov2008essays}
V.~P. Spiridonov, ``{Essays on the Theory of Elliptic Hypergeometric
  Functions},'' {\em Russian Mathematical Surveys} {\bf 63} (2008), no.~3, 405,
  \href{http://www.arXiv.org/abs/0805.3135}{{\tt 0805.3135}}.

\bibitem{spiridonov2005classical}
V.~Spiridonov, ``{Classical Elliptic Hypergeometric Functions and Their
  Applications},'' \href{http://www.arXiv.org/abs/math/0511579}{{\tt
  math/0511579}}.

\bibitem{Witten:1993jg}
E.~Witten, ``{On the Landau-Ginzburg description of N=2 minimal models},'' {\em
  Int. J. Mod. Phys.} {\bf A9} (1994) 4783--4800,
\href{http://www.arXiv.org/abs/hep-th/9304026}{{\tt hep-th/9304026}}.

\end{thebibliography}\endgroup
\bibliographystyle{utphys}

\end{document}